\documentclass[english,prx,aps,article,twocolumn,showpacs,reprint,microtype,longbibliography,superscriptaddress]{revtex4-2}
\usepackage{amsmath}
\usepackage{graphicx}% Include figure files
\usepackage{subfig} % subfloats as later one is not working for subfloats
\usepackage{dcolumn}% Align table columns on decimal point
\usepackage{multirow}
\usepackage{textcomp}
\usepackage{threeparttable} %for footnotes in table
\usepackage{soul} %for highlighting the text
\usepackage{verbatim} %for commenting out whole paragraph by using \begin{comment} and \end{comment}
\usepackage{braket}
\usepackage{color}
\usepackage{xcolor}
\usepackage[unicode=true,pdfusetitle,
bookmarks=true,bookmarksnumbered=false,bookmarksopen=false,
breaklinks=false,pdfborder={0 0 1},backref=false,colorlinks=true]
{hyperref}
\usepackage{siunitx}
\usepackage{graphicx}% Include figure files
\usepackage{dcolumn}% Align table columns on decimal point
\usepackage{bm}% bold math
\begin{document}

\preprint{APS/123-QED}

\title{Upscaling from ab initio atomistic simulations to electrode scale:\\ The case of manganese hexacyanoferrate, a cathode material for Na-ion batteries}% Force line breaks with \\

\author{Yuan-Chi Yang}
\affiliation{Univ. Grenoble Alpes, CEA, Liten, DEHT, 38000 Grenoble, France}

\author{Eric Woillez}
\affiliation{Univ. Grenoble Alpes, CEA, Liten, DEHT, 38000 Grenoble, France}

\author{Quentin Jacquet}
\affiliation{Univ. Grenoble Alpes, CEA, CNRS, Grenoble INP, IRIG, SyMMES, 38000 Grenoble, France}

\author{Ambroise Van Roekeghem}
\email{ambroise.vanroekeghem@cea.fr}
\affiliation{Univ. Grenoble Alpes, CEA, Liten, DEHT, 38000 Grenoble, France}


\begin{abstract}

We present a generalizable scale-bridging computational framework that enables predictive modeling of insertion-type electrode materials from atomistic to device scales. Applied to sodium manganese hexacyanoferrate, a promising cathode material for grid-scale sodium-ion batteries, our methodology employs an active-learning strategy to train a Moment Tensor Potential through iterative hybrid grand-canonical Monte Carlo–-molecular dynamics sampling, robustly capturing configuration spaces at all sodiation levels. The resulting machine learning interatomic potential accurately reproduces experimental properties including volume expansion, operating voltage, and sodium concentration-dependent structural transformations, while revealing a four-order-of-magnitude difference in sodium diffusivity between the rhombohedral (sodium-rich) and tetragonal (sodium-poor) phases at 300~K. We directly compute all critical parameters—-temperature- and concentration-dependent diffusivities, interfacial and strain energies, and complete free-energy landscapes-—to feed them into pseudo-2D phase-field simulations that predict phase-boundary propagation and rate-dependent performances across electrode length scales. This multiscale workflow establishes a blueprint for rational computational design of next-generation insertion-type materials, such as battery electrode materials, demonstrating how atomistic insights can be systematically translated into continuum-scale predictions.
\end{abstract}

\maketitle 
%\tableofcontents

\section{\label{sec:level1}Introduction}
Sodium manganese hexacyanoferrate ($\text{MnFePBA}$) has emerged as one of the most promising cathode materials for sodium-ion batteries. Its earth-abundant manganese and iron redox centers, low-cost synthesis, and high reversible capacity (120--140 $\text{mAh g}^{-1}$) \cite{ref1, ref2} make it particularly attractive for grid-scale energy storage applications, where both performance and economic viability are critical. 

However, designing high-performance batteries remains challenging due to the complexity of battery operation, which involves coupled charge transfer and mass transport across multiple length scales---from crystal lattices, to individual particles, and finally to complete electrodes. Conventional design optimization relies on iterative trial-and-error approaches that are both time-intensive and resource-demanding. A more efficient strategy would employ computational digital twins capable of predicting electrochemical and physical properties across multiple scales, enabling optimization of critical parameters -- such as particle size, electrode thickness, porosity, and charging protocols -- through simulation rather than experiment.

Current computational studies of battery electrode materials typically focus on isolated length scales: density functional theory ($\text{DFT}$) and molecular dynamics ($\text{MD}$) for atomic-scale phenomena \cite{ref3, ref4, ref5, ref6}, phase-field models for particle-scale behavior \cite{ref7}, and pseudo-2D ($\text{P}2\text{D}$) models for electrode-scale processes \cite{ref8}. Recognizing the importance of bridging these disparate scales, recent efforts have attempted to construct complete phase diagrams by linking atomistic energetics to continuum thermodynamics. The prevailing approach involves using either cluster expansion ($\text{CE}$) methods or machine learning interatomic potentials ($\text{MLIPs}$) trained on $\text{DFT}$ datasets to compute formation energies of numerous configurations generated through semi-grand canonical Monte Carlo (semi-$\text{GCMC}$) simulations, from which phase diagrams and voltage profiles are derived via convex hull analysis.

However, both methodologies exhibit significant limitations that compromise their predictive accuracy. $\text{MLIPs}$ face critical challenges in dataset coverage and quality. For instance, Kwon and Kim demonstrated that graph neural network-based $\text{MLIPs}$ trained on structures from the Materials Project incorrectly predicted monophasic behavior in $\text{Li}_{1-x}\text{FePO}_4$ ($\text{LFP}$), contradicting the well-established biphasic reaction observed experimentally and in $\text{DFT}$ calculations \cite{ref9}. This limited predictive capability resulted from incomplete coverage of the full compositional space during training, highlighting that $\text{MLIP}$ accuracy depends critically on comprehensive training datasets that span all relevant configurations.

Cluster expansion ($\text{CE}$) methods face complementary but equally fundamental limitations. Unless vibrational free-energy corrections or ``cluster-plus-strain'' terms are explicitly included, $\text{CE}$ effectively operates as a ground-state configurational Hamiltonian, neglecting strain effects and vibrational entropy. This leads to systematic discrepancies between predicted and experimental voltage profiles across diverse battery systems, including disordered rocksalt $\text{Li}_3\text{V}_2\text{O}_5$ \cite{ref10}, $\text{Li}_x\text{NiO}_2$ \cite{ref11}, and $\text{Li}_x\text{CoO}_2$ \cite{ref12}. The rigid-lattice assumption inherent to $\text{CE}$ proves particularly problematic for systems undergoing structural phase transitions or exhibiting large local distortions. For example, $\text{Li}_x\text{CoO}_2$ undergoes stacking sequence transitions from $\text{O}3$ to $\text{H}1-3$ structure below $x = 1/3$, limiting the validity of $\text{O}3$-based $\text{CE}$ predictions to compositions $x > 1/3$.

Furthermore, constructing complete phase diagrams through $\text{CE}$ requires a labor-intensive multi-step process: (i) determining the number of equilibrium phases at different concentrations; (ii) training separate $\text{CE}$ models for each structural phase; and (iii) comparing convex hull predictions from each model. Systems experiencing drastic structural changes, such as the rhombohedral-to-tetragonal transition in $\text{MnFePBA}$ during sodium insertion and removal, remain particularly challenging to describe accurately. These limitations collectively highlight the need for more sophisticated computational frameworks capable of capturing the coupled thermodynamic, kinetic, and structural evolution across multiple length scales in battery materials.

In this article, we present a generalizable scale-bridging workflow to construct a $\text{P}2\text{D}$ phase-field model of the electrode, demonstrating predictive modeling of insertion-type electrode materials from atomistic to device scales.

The critical bottleneck in $\text{P}2\text{D}$ phase-field modeling has been the accurate determination of concentration-dependent properties, including free-energy landscapes, diffusivities (varying with both temperature and concentration), and interfacial energies. These properties have traditionally been intractable to measure experimentally or compute using conventional simulation techniques such as $\text{DFT}$, ab initio molecular dynamics ($\text{AIMD}$), or classical molecular dynamics ($\text{MD}$) alone \cite{ref13}. The challenge stems from the expensive computational cost required to accurately (near ab initio accuracy) simulate large system sizes ($\approx 10^{3} \text{--} 10^{4}$ atoms) over long timescales ($\approx 10^{-9}$ s) necessary to adequately sample these properties while maintaining high accuracy.

We overcome this limitation using an active learning workflow to train a Moment Tensor Potential ($\text{MTP}$) \cite{ref14}, which is a machine-learning interatomic potential tailored here for $\text{MnFePBA}$. Our selection of $\text{MTP}$ is motivated by its superior performance: firstly, $\text{MTP}$ models demonstrate superior prediction accuracy, computational efficiency, and data efficiency compared to other $\text{MLIP}$ models \cite{ref15}. Furthermore, $\text{GCMC}$ sampling requires intensive energy evaluations for structures involving random insertion, deletion, and translation moves that often lie outside the training set. The $\text{MTP}$ model, which predicts interatomic forces and atomic energies through a linear combination of non-linear basis functions, exhibits reliable extrapolation and interpolation of forces and energies \cite{ref17}, allowing for long-term $\text{GCMC-MD}$ simulation.

Our active-learning strategy iteratively enriches the training dataset by combining machine learning force field ($\text{MLFF}$) assisted $\text{AIMD}$ \cite{ref18, ref19} at the pre-training stage, and hybrid grand-canonical Monte Carlo--molecular dynamics ($\text{GCMC-MD}$) sampling \cite{ref20} during the $\text{GCMC}$ process itself. This approach ensures that rare, high-energy configurations sampled during $\text{GCMC}$ are incorporated into the training set, yielding a robust $\text{MLIP}$ trained on 3,593 configurations spanning the full sodium concentration range and diverse coordination environments.

The resulting $\text{MLIP}$ accurately reproduces experimentally measured properties, including volume expansion upon desodiation, average operating voltage, and local structural distortions. Leveraging this validated $\text{MLIP}$, we perform large-scale molecular dynamics simulations to compute essential inputs for continuum modeling. Specifically, we extract: (1) temperature- and concentration-dependent $\text{Na}^{+}$ diffusivities and activation energies over nanosecond timescales, (2) interfacial and strain energies, and (3) free-energy profiles for $\text{Na}$ insertion via biased $\text{GCMC}$ sampling. These atomistic properties feed directly into $\text{P}2\text{D}$ phase-field simulations, enabling predictive modeling of phase-boundary propagation, particle-level and electrode-level state-of-charge heterogeneity, and rate-dependent electrochemical response.

This scale-bridging framework not only accurately captures the thermodynamics and kinetics of $\text{MnFePBA}$ but also establishes a blueprint for  the  accelerated  computational  design  and optimization of next-generation insertion-type electrode materials, by accurately predicting battery performance across multiple length scales. The remainder of this paper is organized as follows: we first describe the active-learning workflow and $\text{MLIP}$ training procedure, then present the atomistic properties calculations, followed by $\text{P}2\text{D}$ phase-field model construction and validation, and finally demonstrate predictive simulations of battery performances under various operating conditions.

\begin{figure*}
\includegraphics[width=17cm]{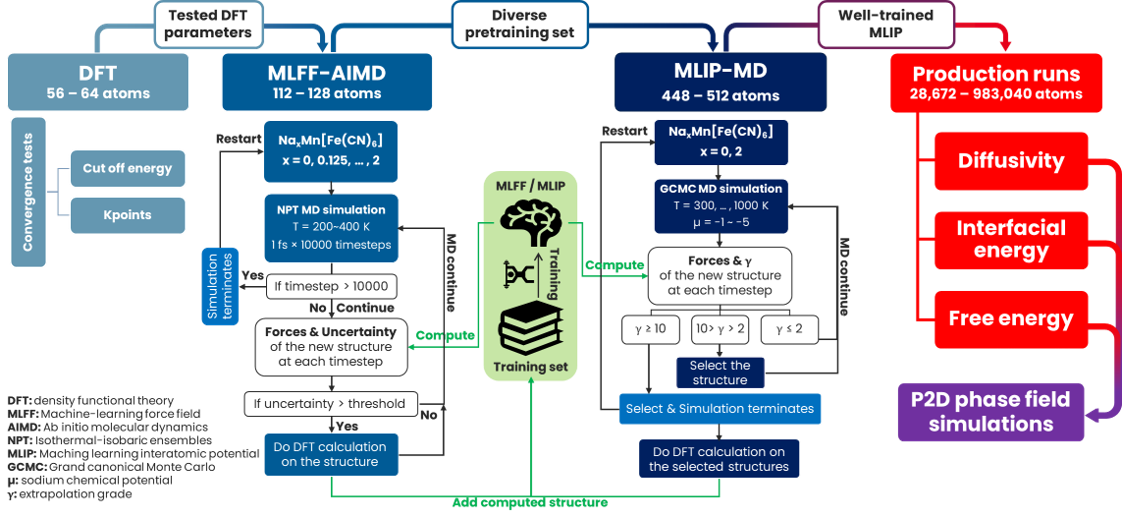}% Here is how to import EPS art
\caption{\label{fig:fig_1}Schematic representation of the three phases in the machine-learning interatomic potential (MLIP) training workflow. Phase one involves performing convergence tests on the density functional theory (DFT) parameters required for subsequent phases. In the second phase, machine-learning force field (MLFF) assisted active learning within \textit{ab initio} molecular dynamics (AIMD) is performed on structures with various sodium concentrations to efficiently acquire a diverse pre-training dataset. In the third phase, the Moment Tensor Potential (MTP) model is first trained on this pre-training dataset. Subsequently, active learning is performed during grand-canonical Monte Carlo--molecular dynamics (GCMC-MD) simulations at different temperatures and chemical potentials for multiple iterations until the MTP no longer selects new structures, which serves as the criterion for convergence. The final trained MTP-MLIP is then utilized to perform production runs on large-scale atomistic systems to calculate the physical properties of MnFePBA.}
\end{figure*}

\section{\label{sec:level1}Computational methods}

%%%%%%%%%%%%%%%%%%%%%%%%%%%%%%%%%%%%%%%%%%%%%%%%%%%%
\subsection{\label{sec:level2}Density functional theory calculations}
$\text{DFT} + \text{U}$ calculations were carried out using the Vienna \textit{ab initio} simulation ($\text{VASP}$) package \cite{ref21, ref22} with the $\text{PBEsol}$ exchange-correlation functional within the projector augmented wave ($\text{PAW}$) method. 

To properly account for the strong on-site Coulomb interactions within the localized $3d$ orbitals of transition metals, $U$ parameters are applied to Fe and Mn ions. A high plane-wave energy cutoff of 700 $\text{eV}$ was employed to accurately describe the short interatomic distances in the system. $\Gamma$-centered $1 \times 1 \times 1$ and $2 \times 2 \times 2$ $\text{k}$-point meshes were tested. We determined that a single $\Gamma$-point is sufficient due to the large size of the supercells used for $\text{AIMD}$ and for the active-learning selected structures during $\text{GCMC-MD}$. The choice of $U$ values is discussed below. 

Xiao et al. \cite{ref5} initially adopted $U$ values of $4.3 \text{ eV}$ for Fe and $5.0 \text{ eV}$ for Mn also used in olivine materials ($\text{LiFePO}_4$ and $\text{LiMnPO}_4$). However, as these parameters failed to capture the phase diagram of their specific system accurately, they recalibrated the values to $2.0 \text{ eV}$ for Fe and $4.0 \text{ eV}$ for Mn to align with higher-accuracy HSE06 hybrid functional calculations.

Further variations are seen in the work of Ito et al. \cite{ref35, Ito2026}, who derived their parameters from existing literature \cite{hegner2016database}, applying a combination of $5.0 \text{ eV}$ for Fe and $4.0 \text{ eV}$ for Mn. Recent studies highlight the sensitivity of these parameters to the electronic environment. Baumgart et al. \cite{Baumgart2026} and Wang et al. \cite{Li2025} utilized distinct $U$ values of $7.0 \text{ eV}$ for high-spin Fe and $3.0 \text{ eV}$ for low-spin Fe. While these values were adopted from previous research \cite{wojdel2008prediction} demonstrating accurate structural and electronic predictions for hexacyanoferrate compounds, they still significantly underestimate open-circuit voltages (OCVs) and electronic band gaps. We also have to keep in mind that a given value of $U$ is linked to the DFT functional on which the Hubbard Hamiltonian is based on, which can explain this large variability in the different studies.

In this work, we take a $U$ value of $5.5 \text{ eV}$ for both Fe and Mn localized $3d$ orbitals, implemented in conjunction with the PBEsol exchange-correlation functional within the Dudarev scheme. This $U$ value was specifically calibrated to match the experimental bandgap of cubic Prussian Blue $\sim 1.75 \text{ eV}$ \cite{robin1962color}, thereby ensuring that the framework rigidity and electronic properties of the MnFePBA system are accurately captured. Furthermore, this parameterization successfully predicts the rhombohedral phase as the stable configuration upon sodiation. 

All DFT calculations were spin-polarized. We initialized the transition metal ions based on their expected electronic configurations in the MnFePBA framework. Mn ions were initialized in a high-spin (HS) state, and Fe ions were initialized in a low-spin (LS) state. To ensure a consistent baseline across the thousands of configurations sampled during the active-learning process, we utilized a ferromagnetic initial ordering.

During the electronic optimization process, the magnetic moments were allowed to relax freely. After numerous electronic steps, the magnetic moment of each Fe and Mn atom converged to its respective ground-state value for that specific atomic configuration. This flexibility allows the model to capture the evolution of oxidation and spin states—such as the mixture of Fe$^{3+}$/Mn$^{2+}$ and Fe$^{3+}$/Mn$^{3+}$—that occurs during the (de)sodiation process.

%%%%%%%%%%%%%%%%%%%%%%%%%%%%%%%%%%%%%%%%%%%%%%%%%%%%
\subsection{\label{sec:level2}Ab initio molecular dynamics (AIMD) with an active learning strategy}
In the first phase, \textit{ab initio} molecular dynamics ($\text{AIMD}$) simulations in the $\text{NPT}$ ensemble (where the number of particles, temperature, and pressure are fixed) were performed on $\text{MnFe-PBA}$ at 16 different sodium concentrations ($\text{Na}_{x}\text{Mn}[\text{Fe}(\text{CN})_{6}]$, $x = 0.125, 0.25, \dots, 2$). Each simulation consisted of 10,000 time steps with a timestep of 1 $\text{fs}$, during which the temperature was ramped from 200 $\text{K}$ to 400 $\text{K}$. Instead of periodic sampling, an active learning strategy guided by a machine-learning force field ($\text{MLFF}$) implemented in $\text{VASP}$ \cite{ref18, ref19} was employed. The $\text{MLFF}$ actively identified and selected structures with force or energy prediction errors exceeding a defined threshold, ensuring the inclusion of the most informative and diverse configurations. The selected structures were then evaluated using $\text{DFT} + \text{U}$ calculations to obtain accurate energies and interatomic forces, which were used to iteratively update and refine the $\text{MLFF}$. This $\text{AIMD}$--$\text{MLFF}$ cycle was repeated for all 16 compositions until the model no longer identified new high-error structures, indicating convergence of the training process. This strategy enabled efficient sampling of the potential energy surface ($\text{PES}$), avoiding redundancy and improving coverage of relevant configurations. The resulting dataset of 677 configurations provided a robust initial foundation for the second-phase training of the machine-learning interatomic potential ($\text{MLIP}$), enhancing both model accuracy and training efficiency.

%%%%%%%%%%%%%%%%%%%%%%%%%%%%%%%%%%%%%%%%%%%%%%%%%%%%
\subsection{\label{sec:level2}MLIP training and hybrid Grand Canonical Monte Carlo (GCMC) -- MD simulations}
In the second phase, hybrid Grand Canonical Monte Carlo--Molecular Dynamics ($\text{GCMC-MD}$) simulations were carried out on $\text{MnFe-PBA}$ under varying sodium chemical potentials to model the (de)sodiation process, utilizing the Large-scale Atomic/Molecular Massively Parallel Simulator ($\text{LAMMPS}$) \cite{ref23}. These simulations involve sodium ion insertion and deletion (grand canonical exchanges), as well as translational Monte Carlo ($\text{MC}$) moves, allowing efficient sampling of partially (de)sodiated structures and further enriching the training dataset. The machine-learning interatomic potential ($\text{MLIP}$) employed is the Moment Tensor Potential ($\text{MTP}$) \cite{ref14, ref24}, which leverages the MaxVol algorithm \cite{ref16} to evaluate the novelty---quantified by the extrapolation grade ($\gamma$)---of each sampled structure relative to those already present in the training set. A level 12 \text{MTP} model constructed using 634 parameters with a cutoff radius of 5~\text{\AA} was used to reach the balance between accuracy and computational cost.

Fully (de)sodiated structures (x=0, 2) were the input structures to perform GCMC-MD at various sodium chemical potentials. At each time step, the $\text{MLIP}$ computes atomic forces and the $\gamma$ value of the newly generated structure. If $2 < \gamma < 10$, the structure is flagged for subsequent $\text{DFT}$ calculations; if $\gamma > 10$, the current $\text{MLIP}$ is deemed unreliable, and the $\text{GCMC-MD}$ simulation is halted. Selected structures are then evaluated using single-point $\text{DFT+U}$ calculations to obtain reference energies, forces, and stress tensors. During the calculation, only the electronic optimization was performed while the atomic positions (ionic coordinates) and cell parameters remain fixed. To prevent contamination of the training set by highly unphysical structures, we impose screening criteria: structures with forces exceeding 150 $\text{eV}/\AA$ or energies above 0 $\text{eV}$ are discarded. The screened $\text{DFT}$ dataset is then used to expand and retrain the $\text{MLIP}$, improving its predictive capability in previously undersampled regions of configuration space. 

The $\text{GCMC-MD}$ protocol proceeds in alternating cycles, where a phase of 4,000 $\text{NPT}$ time steps (with a timestep of 0.4 $\text{fs}$) is followed by 1,000 grand canonical exchange attempts and 500 $\text{MC}$ moves The $\text{NPT}$ simulation phase serves to relax the system following sodium insertion or removal, ensuring structural equilibration and thermodynamic consistency. This hybrid sampling strategy efficiently explores the configurational space. The active learning loop is repeated until MaxVol selects no new configurations, indicating that the $\text{MLIP}$ has achieved sufficient coverage of the relevant phase space.

%%%%%%%%%%%%%%%%%%%%%%%%%%%%%%%%%%%%%%%%%%%%%%%%%%%%
\subsection{\label{sec:level2}Diffusivity and activation energy}
To evaluate the sodium-ion diffusivities and corresponding activation energies, molecular dynamics simulations were conducted using the trained MLIP model on structures generated from GCMC sampling across different sodium concentrations. The simulation cell size was set to $80 \times 80 \times 80$ \AA$^3$, containing approximately 28,000 to 32,000 atoms, depending on the sodium content (0--4096 Na$^+$ ions). Snapshots of the simulated system are shown in Fig.~\ref{fig:fig_S2}. Each simulation began with a $20\text{ ps}$ thermalization phase under the NPT ensemble to ensure structural equilibration. Following thermalization, production NPT simulations were run for up to $5\text{ ns}$, during which the mean square displacement (MSD) of sodium ions was computed.

The MSD was averaged over all Na$^+$ ions in the system and analyzed as a function of time interval $\Delta t$ to extract the diffusion coefficient using the Einstein relation:
\begin{equation}
D = \frac{\text{MSD}(\Delta t)}{2d \Delta t}
\end{equation}
where ($d = 3$) is the spatial dimensionality. The MSD is calculated as:
\begin{equation}
\text{MSD}(\Delta t) = \frac{1}{N} \sum_{i=1}^{N} \langle |\mathbf{r}_i(t + \Delta t) - \mathbf{r}_i(t)|^2 \rangle
\end{equation}
where $\mathbf{r}_i(t)$ is the position of ion $i$ at time $t$, and the brackets denote an average over time origins. By performing these simulations at multiple temperatures, an Arrhenius analysis was conducted to extract the activation energy $E_{\text{a}}$, according to:
\begin{equation}
D(T) = D_0 \exp\left(-\frac{E_{\text{a}}}{kT}\right)
\end{equation}
where $D_0$ is the pre-exponential factor, $k$ is Boltzmann's constant, and $T$ is the absolute temperature.

%%%%%%%%%%%%%%%%%%%%%%%%%%%%%%%%%%%%%%%%%%%%%%%%%%%%
\subsection{\label{sec:level2}Interfacial energy and strain energy}
To evaluate the interfacial and strain energies of the system, we first constructed a series of eight fully sodiated rhombohedral cells with fixed $y$ and $z$ dimensions but varying $x$ dimensions. These cells were generated by duplicating an $80 \times 80 \times 80$ \AA$^3$ fully sodiated rhombohedral unit cell along the $x$ direction from one to 32 times. For each resulting supercell, all sodium atoms in the upper half along the $x$ direction were removed to create two interfaces under periodic boundary conditions. The resulting interfacial systems were then relaxed by performing 20~ps of NPT MD simulations. Finally, the potential energy of each relaxed cell was calculated using the trained MTP-MLIP model.

The interfacial and strain energies were determined by evaluating the excess energy of the interfacial systems arising from phase separation. The excess energy was calculated as the difference between the total energy of the two-phase system and the sum of the energies of the corresponding pure phases, as expressed in Eqs.~(\ref{eq:eq_4}) \cite{ref25}:

\begin{equation}
E_{\text{excess}} = E_{N \text{ unit cells}}^{\text{2-phase}} - \frac{N}{2} E_{\text{Na}_2\text{Mn}[\text{Fe(CN)}_6]} - \frac{N}{2} E_{\text{Mn}[\text{Fe(CN)}_6]}
\label{eq:eq_4}
\end{equation}

This excess energy consists of two contributions: the interfacial energy and the strain energy, which can be related through Equation~(\ref{eq:eq_5}):

\begin{equation}
E_{\text{excess}} = 2A_{\text{interface}}\gamma + V_{\text{particle}}e_{\text{strain}}
\label{eq:eq_5}
\end{equation}

where $A_{\text{interface}}$ is the interfacial area, $\gamma$ is the interfacial energy density, $V_{\text{particle}}$ is the particle volume, and $e_{\text{strain}}$ is the strain energy density.
%%%%%%%%%%%%%%%%%%%%%%%%%%%%%%%%%%%%%%%%%%%%%%%%%%%%
\subsection{\label{sec:level2}Bias potential free energy sampling}

To explore the configurational space of sodium intercalation in Prussian blue analogs, we employ a bias-enhanced hybrid ($\text{GCMC-MD}$) method \cite{ref26}. This approach enables direct sampling of configurations and chemical potential ($\mu_{\text{Na}}$), while systematically encouraging exploration around a target sodium content.The chemical potential of sodium is defined thermodynamically as the partial derivative of the Gibbs free energy with respect to sodium content at constant pressure and temperature:

\begin{equation}
\mu(N_{\text{Na}}) = \left( \frac{\partial G}{\partial N_{\text{Na}}} \right)_{P, T}
\label{eq:eq_6}
\end{equation}

To improve sampling over different sodium concentrations, we introduce a harmonic bias potential centered at a target sodium number \cite{ref26}:

\begin{equation}
B(N_{\text{Na}}) = \phi(N_{\text{Na}} - N_{\text{Na}}^t)^2
\label{eq:eq_7}
\end{equation}

where $N_{\text{Na}}^t$ is the chosen target and $\phi$ controls the strength of the bias; the total biased free energy is then:

\begin{equation}
G_{\text{bias}}(N_{\text{Na}}) = G(N_{\text{Na}}) + B(N_{\text{Na}})
\label{eq:eq_8}
\end{equation}

The partition function of the biased ensemble becomes:

\begin{align}
Z_{\text{bias}}
&= \sum_{\vec{\sigma}}
\exp\Bigl[
-\beta \bigl(
E(\vec{\sigma})
\nonumber \\
&\qquad
- \mu_{\text{bias}} N_{\mathrm{Na}}(\vec{\sigma})
+ B\!\left( N_{\mathrm{Na}}(\vec{\sigma}) \right)
\bigr)
\Bigr]
\label{eq:eq_9}
\end{align}

Here, ($\vec{\sigma}$) denotes a microstate, ($E(\vec{\sigma})$) its total energy of state $\vec{\sigma}$, and $\mu_{\text{bias}}$ is the externally imposed chemical potential for sodium. The symbol $\beta = 1/(k_{\text{B}}T)$ is the inverse thermal energy.

By grouping microscopic configurations by their sodium content, we can rewrite the biased partition function as follows:

\begin{eqnarray} 
Z_{\text{bias}} &&= \sum_{N_{\text{Na}}} \sum_{\vec{\sigma}(N_{\text{Na}})} \exp(-\beta E(\vec{\sigma})) \nonumber\\ &&\exp \left[ -\beta (-\mu_{\text{bias}} N_{\text{Na}} + \phi(N_{\text{Na}} - N_{\text{Na}}^t)^2) \right]
\label{eq:eq_10} 
\end{eqnarray}

where the inner sum defines:

\begin{equation}
Z_{N_{\text{Na}}} = \sum_{\vec{\sigma}(N_{\text{Na}})} \exp(-\beta E(\vec{\sigma})) 
\label{eq:eq_11} 
\end{equation}

with corresponding Helmholtz free energy:

\begin{equation} 
F(N_{\text{Na}}) = -k_{\text{B}}T \ln Z_{N_{\text{Na}}} = G(N_{\text{Na}}) - PV 
\label{eq:eq_12} 
\end{equation}

This leads to:

\begin{equation} 
Z_{N_{\text{Na}}} = \exp[-\beta(G(N_{\text{Na}}) - PV)] 
\label{eq:eq_13} 
\end{equation}

Hence, the total biased partition function is:

\begin{align}
Z_{\text{bias}}
&= \sum_{N_{\text{Na}}}
\exp\Bigl[
-\beta \bigl(
G(N_{\text{Na}})
- \langle PV \rangle_{T,P, N_{\text{Na}}}
\nonumber \\
&\qquad
- \mu_{\text{bias}} N_{\text{Na}}
+ B(N_{\text{Na}})
\bigr)
\Bigr]
\label{eq:eq_14}
\end{align}

From this, the probability of observing a given sodium number in the biased ensemble is:

\begin{align}
P_{\text{bias}}(N_{\text{Na}})
&= \frac{1}{Z_{\text{bias}}}
\exp\Bigl[
-\beta \bigl(
G(N_{\text{Na}})
- \langle PV \rangle_{T,P, N_{\text{Na}}}
\nonumber \\
&\qquad
- \mu_{\text{bias}} N_{\text{Na}}
+ \phi (N_{\text{Na}} - N_{\text{Na}}^{t})^{2}
\bigr)
\Bigr]
\label{eq:eq_15}
\end{align}

At the thermodynamic limit, the average Na content corresponds to the most probable sodium content, which is the maximum of this distribution. Setting the derivative to zero yields:

\begin{align}
&\left.
\frac{\partial P_{\text{bias}}}{\partial N_{\text{Na}}}
\right|_{T,P,N_{\text{Na}}=\langle N_{\text{Na}} \rangle}
=
-\beta P_{\text{bias}}(N_{\text{Na}})
\Bigl[
\mu(N_{\text{Na}})
- \mu_{\text{bias}}
\nonumber \\
&\qquad
- P
\left.
\frac{\partial V}{\partial N_{\text{Na}}}
\right|_{T,P,N_{\text{Na}}=\langle N_{\text{Na}} \rangle}
+ 2\phi \bigl( N_{\text{Na}} - N_{\text{Na}}^{t} \bigr)
\Bigr]
= 0
\label{eq:eq_16}
\end{align}

Solving this condition, we obtain:

\begin{eqnarray} 
\mu(\langle N_{\mathrm{Na}} \rangle) =&& \mu_{\text{bias}} + \left. P \frac{\partial V}{\partial N_{\mathrm{Na}}} \right|_{T, P, N_{\mathrm{Na}} = \langle N_{\mathrm{Na}} \rangle} \nonumber\\ &&
- 2\phi(\langle N_{\mathrm{Na}} \rangle - N_{\mathrm{Na}}^t)
\label{eq:eq_17}
\end{eqnarray} 

By scanning over multiple bias centers $N_{\mathrm{Na}}^t$, we reconstruct the chemical potential function $\mu(N_{\mathrm{Na}})$, and integrate to obtain the free energy \cite{ref27}:

\begin{align}
G(N_{\mathrm{Na}},T)
&= G(N_{\mathrm{Na}}^{0},T)
+ \int_{N_{\mathrm{Na}}^{0}}^{N_{\mathrm{Na}}}
\mu(N_{\mathrm{Na}},T)\,\mathrm{d}N_{\mathrm{Na}}
\label{eq:eq_18}
\end{align}
%%%%%%%%%%%%%%%%%%%%%%%%%%%%%%%%%%%%%%%%%%%%%%%%%%%%
\subsection{\label{sec:level2}P2D phase field simulation}

The thermodynamical quantities related to the MnFePBA cathode material, computed via $\text{GCMC-MD}$ simulations, are integrated into a porous electrode model. There are several macroscopic models of porous electrodes. The choice primarily depends on the level of complexity of the physico-chemical phenomena that the model aims to reproduce \cite{ref36, ref37}. The Doyle-Fuller-Newman model, also referred to as the pseudo two-dimensional ($\text{P2D}$) model, is currently the most widely utilized physics-based model for batteries \cite{ref28, ref29}. It can be considered as the simplest homogeneous model of porous electrodes that includes the main transport mechanisms in the electrolyte and in the solid phase without further approximations than the mathematical limit of volume averaging. Simpler macroscopic models, such as the single particle models with or without electrolyte, are based on a decoupling approximation between the electrolyte and the solid phase. In particular, the single particle model reduces the solid domain to a unique representative particle and cannot therefore represent the complex interparticles dynamics that is emerging from heterogeneous electrochemical activity. The P2D model is thus preferred to simulate a more realistic electrode dynamics.

The model is founded on a volume-averaged representation of various transport phenomena occurring in the electrolyte and within the active material of porous electrodes. Averaged mass and charge transport in the electrolyte is represented in the dimension transverse to the electrode plane. At each position in the electrode depth dimension, a representative active particle is selected, and sodium transport is modeled in the dimension of the particle depth, assuming spherical particle symmetry. 

\begin{figure}
\includegraphics[width=9cm, trim=2 2 2 2, clip]{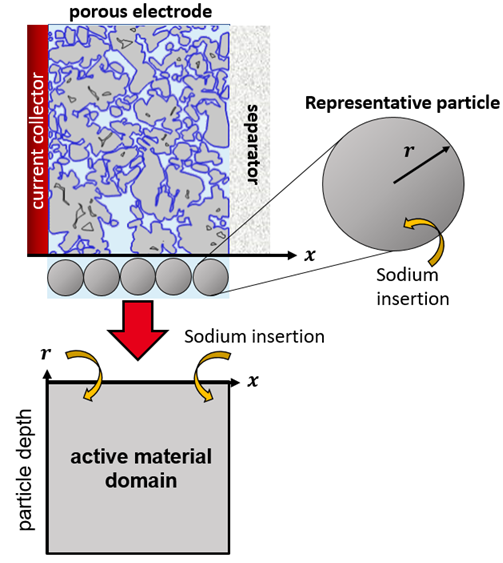}% Here is how to import EPS art
\caption{\label{fig:fig_2}Schematic representation of the pseudo two-dimensional domain of active material in the porous electrode model of Doyle-Fuller-Newman. A porous electrode with current collector and separator is displayed on top. A spherical particle of average particle size is placed at each position in the electrode depth to model sodium transport inside the active material. The pseudo two-dimensional domain (bottom) with the horizontal dimension being the electrode width and the vertical dimension being the particles average radius is the continuous limit with an infinite number of identical representative particles.}
\end{figure}

Consequently, sodium transport in the active material is solved in a pseudo two-dimensional domain where the horizontal dimension ($x$) represents the electrode depth and the vertical dimension ($r$) represents the depth of a representative particle located at position $x$ in the electrode. The active material domain utilized in the $\text{P2D}$ model is illustrated in Fig.~\ref{fig:fig_2}.

Standard $\text{P2D}$ models represent the transport kinetics in the active material by a simple diffusion process based on Fick’s law, where the solid diffusion coefficient can depend on the local sodium concentration. Such transport equations are inadequate for active materials that present phase transitions as a function of the concentration of active species, such as Prussian Blue. Although many active battery materials in both $\text{Li-ion}$ and $\text{Na-ion}$ technology exhibit phase transitions (e.g., graphite, $\text{LFP}$, and lithium-titanium oxide in $\text{Li-ion}$ batteries), phase-field equations are rarely implemented in $\text{P2D}$ models. Their impact on cell performance remains an active research area \cite{ref8, ref30, ref31, ref32}. In the following, we present the phase-field model utilized for Prussian Blue in the $\text{P2D}$ simulations and relate this macroscopic model to the predictions of $\text{GCMC-MD}$ simulations.

The phase-field model is based on the minimization of the total Gibbs free energy:

\begin{equation}
\mathcal{G} = \int_{V} \left( \frac{k}{2} |\nabla N_{\text{Na}}|^2 + G(N_{\text{Na}}) \right) d^3r
\label{eq:eq_19}
\end{equation}

where $G$ is the bulk free energy given by Eqs.~(\ref{eq:eq_18}), and $\frac{k}{2}|\nabla N_{\text{Na}}|^2$ is a penalty term accounting for the excess free energy in the presence of an interface. Integration is performed over the particle volume $V$. The minimization is conducted under the constraint that the total sodium mass is conserved. The equation for mass conservation is given by:

\begin{equation}
\frac{\partial N_{\text{Na}}}{\partial t} = -\nabla \cdot J_{\text{Na}}
\label{eq:eq_20}
\end{equation}

and the sodium flux density is expressed as:

\begin{equation}
J_{\text{Na}} = -M \nabla \frac{\delta \mathcal{G}}{\delta N_{\text{Na}}} = -M \nabla (-k \Delta N_{\text{Na}} + \mu(N_{\text{Na}}))
\label{eq:eq_21}
\end{equation}

In Eq.~(\ref{eq:eq_21}), $M$ is the mobility, and $\mu = \frac{\delta \mathcal{G}}{\delta N_{\text{Na}}}$ is the chemical potential given by Eq.~(\ref{eq:eq_6}). The combination of Eq.~(\ref{eq:eq_20}) and Eq.~(\ref{eq:eq_22}) gives the standard Cahn-Hilliard equation used for conservative phase-transition models:

\begin{equation}
\frac{\partial N_{\text{Na}}}{\partial t} = \nabla \cdot M \nabla (-k \Delta N_{\text{Na}} + \mu(N_{\text{Na}}))
\label{eq:eq_22}
\end{equation}
\\

The Cahn-Hilliard equation requires as input the value of the penalty coefficient $k$, the value of the mobility $M$, and the chemical potential $\mu$ as a function of the sodium concentration $N_{\text{Na}}$. We now explain how these three parameters can be deduced from the physical quantities predicted by Monte-Carlo simulations.

The chemical potential is given by Eq.~(\ref{eq:eq_6}), and can thus be directly implemented in the phase-field model. Importantly, the model gives a phase transition only if the chemical potential is decreasing in some concentration range. The local maximum and minimum give the concentrations of the spinodal decomposition, and the construction of the Maxwell plateau gives the concentrations of the binodal points. The chemical potential of the MnFePBA is represented for different temperatures in Fig.~\ref{fig:fig_8}.

The mobility can have different values in the two phases. Considering Eq.~(\ref{eq:eq_22}) without an interface, we obtain the non-linear Fick equation:

\begin{equation}
\frac{\partial N_{\text{Na}}}{\partial t} = \nabla \cdot M \frac{\partial \mu}{\partial N_{\text{Na}}} \nabla N_{\text{Na}}
\label{eq:eq_23}
\end{equation}

The solid diffusion coefficient is thus related to the phase-field parameters by:

\begin{equation}
D_s = M \frac{\partial \mu}{\partial N_{\text{Na}}}
\label{eq:eq_24}
\end{equation}

The values for the solid diffusion coefficient in each phase at different temperatures are displayed in Fig.~\ref{fig:fig_6}.

The last step is to compute the value of $k$ using its relation to the interfacial energy density $\gamma$. This can be done by the computation of the excess free energy at equilibrium, due to the presence of a non-vanishing interface.

\begin{widetext}
\begin{equation}
\Delta \mathcal{G} = \int_{V} \frac{k}{2} |\nabla N_{\text{Na}}^{eq}|^2 + G(N_{\text{Na}}^{eq}) \, d^3r - \int_{V_1} G(N_{\text{Na}}^1) \, d^3r - \int_{V_2} G(N_{\text{Na}}^2) \, d^3r
\label{eq:eq_25}
\end{equation}
\end{widetext}

where $N_{\text{Na}}^{eq}(r)$ is the equilibrium concentration profile in an infinite domain $\mathcal{V}$. The bulk concentrations of the two phases, given by the binodal points, are $N_{\text{Na}}^1$ and $N_{\text{Na}}^2$, respectively. We obtain the expression:

\begin{equation}
\gamma = \sqrt{2k} \int_{N_{\text{Na}}^1}^{N_{\text{Na}}^2} \sqrt{p(N)} \, dN
\label{eq:eq_26}
\end{equation}

where the function $p(N)$ is defined as:

\begin{equation}
p(N) = G(N) - G(N_{\text{Na}}^1) - \mu(N_{\text{Na}}^1) \cdot (N - N_{\text{Na}}^1)
\label{eq:eq_27}
\end{equation}

The typical width $l^{int}$ of the interface scales as:

\begin{equation}
l^{int} \sim \frac{1}{N_{\text{Na}}^2 - N_{\text{Na}}^1} \sqrt{\frac{k}{2 \max(p)}}
\label{eq:eq_28}
\end{equation}

In the standard P2D model, the insertion/desinsertion kinetics in the active material is given by the Butler-Volmer law:

\begin{equation}
j_{out} = j_0 \sqrt{\frac{c_l}{c_0}} \left[ e^{\frac{\beta \eta}{2}} - e^{-\frac{\beta \eta}{2}} \right]
\label{eq:eq_29}
\end{equation}

where $j_{out}$ is the current density, $j_0$ is a kinetic parameter, $c_l$ is the electrolyte concentration at the surface, and $c_0$ is the average electrolyte concentration. The coefficient $\beta = 1/k_B T$ is the inverse thermal energy. The overpotential $\eta$ is defined as:

\begin{equation}
\eta = e\phi - e\phi_l + \frac{\delta \mathcal{G}}{\delta N_{\text{Na}}}
\label{eq:eq_30}
\end{equation}

where $e$ is the electron charge, $\phi$ is the material electric potential, $\phi_l$ is the electrochemical potential of sodium ions, and $\frac{\delta \mathcal{G}}{\delta N_a}$ is the thermodynamic force presented in Eq.~(\ref{eq:eq_20}). When the concentration field at the particle surface is homogeneous, $\frac{\delta \mathcal{G}}{\delta N_a} = \mu$ and Eq.~(\ref{eq:eq_30}) reduces to the standard Buttler-Volmer law.

\section{\label{sec:level1}Results and discussion}
%%%%%%%%%%%%%%%%%%%%%%%%%%%%%%%%%%%%%%%%%%%%%%%
\subsection{\label{sec:level2}Validation of MLIP}
\begin{figure*}
\includegraphics[width=14cm, trim=2 2 2 2, clip]{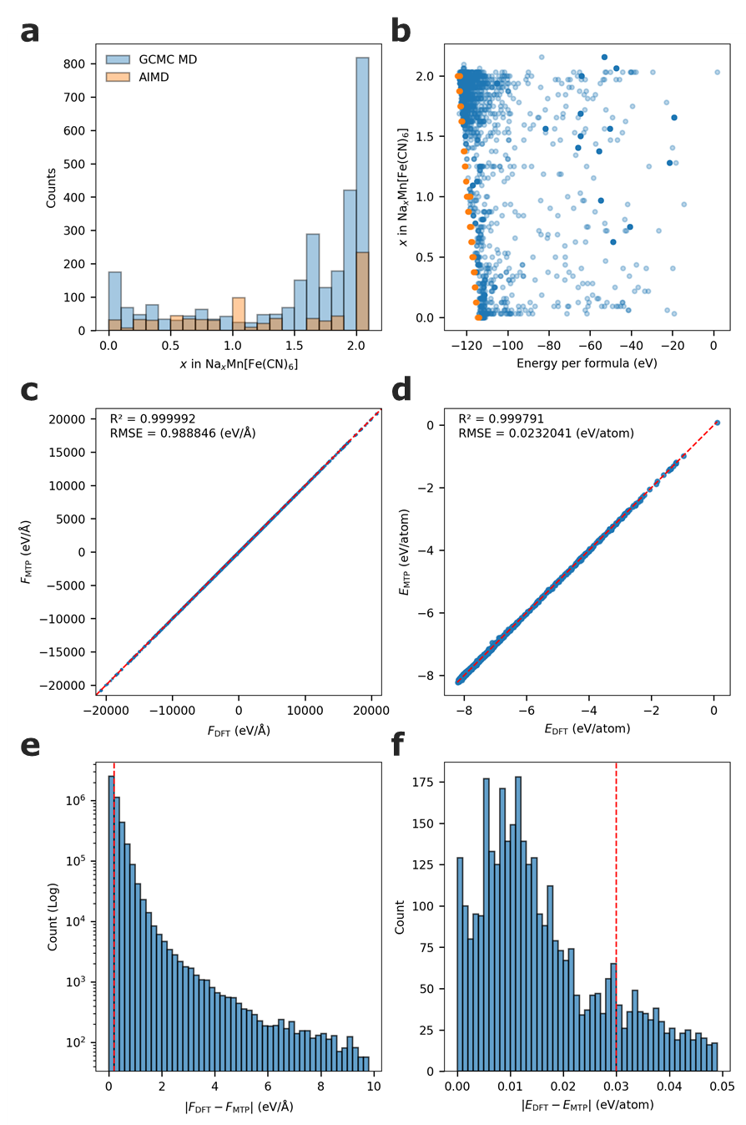}% Here is how to import EPS art
\caption{\label{fig:fig_3}(a) presents the distribution of configurations as a histogram plotted against sodium concentration, where the pre-training data (orange) from ab initio molecular dynamics (AIMD) is distinguished from the GCMC-MD generated structures, which show dominant sampling at $x = 0$ and $x = 2$. (b) illustrates the wide energy dispersion of the training structures. (c) and (d) show the scatter plots comparing the MTP-predicted forces and energies, respectively, against the DFT-calculated reference values. (e) and (f) show histograms of the error distributions for forces and energies, respectively: over 81\% of force errors fall within 0.2~eV/\AA, while over 81\% of energy errors per atom remain within 0.03~eV/atom. The vertical red dashed lines indicate the 0.2~eV/\AA, and 0.03~eV/atom thresholds.}
\end{figure*}

The final MTP training set comprises 3,593 structures containing a total of 1,500,221 atoms, sampled across the full range of sodium concentrations. Fig.~\ref{fig:fig_3}a presents a histogram of the configuration distribution as a function of sodium content. Pre-training data and configurations generated from AIMD and GCMC-MD are shown in orange and blue, respectively. The training set provides comprehensive coverage across all sodium concentrations, with higher sampling density at $x = 2$ and $x = 0$ because the initial structures for GCMC-MD active learning were either fully sodiated or fully desodiated. 

Fig.~\ref{fig:fig_3}b shows the energy distribution of structures in the training, demonstrating the wide energy dispersion that reflects the structural diversity captured during sampling. It is evident that NPT AIMD alone is insufficient to sample the complete phase space. However, it provides the MTP model with essential atomistic environmental information, enabling more efficient sampling during GCMC-MD. Fig.~\ref{fig:fig_3}c and \ref{fig:fig_3}d present scatter plots comparing MTP-predicted versus DFT-calculated forces and energies, respectively. The forces and energies prediction error histograms are shown in Fig.~\ref{fig:fig_3}e and Fig.~\ref{fig:fig_3}f. Over 81\% of force errors fall within 0.2~eV/\AA, while over 81\% of energy errors per atom remain within 0.03~eV/atom. This level of accuracy is within the acceptable range for capturing the lattice vibrations and ionic diffusion behavior of Prussian Blue analogs at the temperatures studied.

Despite having elevated root mean square errors (RMSEs) of 0.989~eV/\AA\ for forces and 0.0232~eV/atom for energies—values that exceed those typically reported for well-converged MLIPs trained on equilibrium structures (RMSE energy error: $\sim$0.01~meV/atom, RMSE force error: $\sim$0.20~eV/\AA\ \cite{ref15})—the trained MTP demonstrates robust transferability. It captures the entire range of atomistic environments, with forces spanning from 0 up to 150~eV/\AA, significantly exceeding the force range (usually up to 10~eV/\AA) reported in other MLIP studies \cite{ref33, ref34}. 

This elevated error arises from the inherent diversity and complexity of the training data, which includes configurations generated via GCMC-MD simulations. During GCMC sampling, random sodium insertion and translation moves occasionally produce unphysical configurations with extremely short interatomic distances, resulting in large forces and energies. However, incorporating these high-energy configurations is essential: they ensure accurate evaluation of GCMC acceptance probabilities and enable stable, reliable simulations across the entire sodium concentration range. Without these configurations, the MTP would lack the necessary transferability to handle the rare but important structural excursions encountered during grand-canonical sampling.

A separate validation set, composed of 70 structures with energies ranging from $-2000$~eV to above $4000$~eV and interatomic forces exceeding 150~eV/\AA, was employed to test the extrapolation capability of the trained MTP (Fig.~\ref{fig:fig_S3}a). The trained MTP also reliably extrapolates and accurately predicts these forces and energies, yielding $R^2$ values of 0.998 and 0.999 for forces and energies, respectively. This outcome confirms the robustness of the trained MTP and the effectiveness of the active learning strategy.

%%%%%%%%%%%%%%%%%%%%%%%%%%%%%%%%%%%%%%%%%%%%%%%
\subsection{\label{sec:level2}Physical properties}

\begin{figure*}
\includegraphics[trim=2 2 2 2, clip]{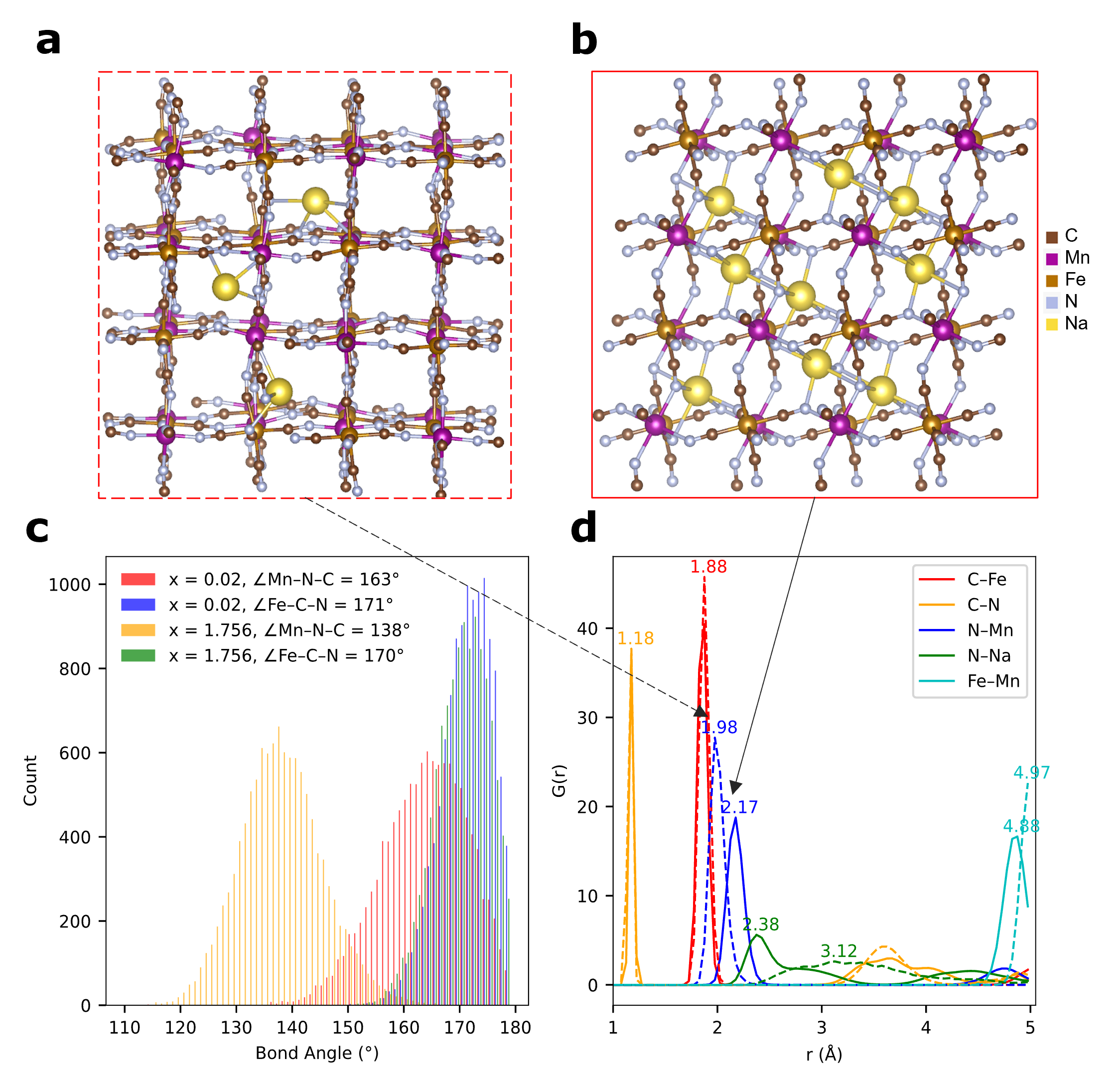}% Here is how to import EPS art
\caption{\label{fig:fig_4}The crystal structure of the MnFePBA at (a) x = 0.02 in tetragonal symmetry and (b) x = 1.756 in rhombohedral symmetry. (c) Bond angle distribution for the sodium-poor (x=0.02) and sodium-rich (x=1.756) structures at 300 K. (d) Radial distribution functions (RDFs) for various atom pairs in the sodium-poor (dashed lines) and sodium-rich (solid lines) structures at 300 K.}
\end{figure*}

Despite the elevated root mean square errors (RMSEs) in training, the developed Machine Learning Interatomic Potential (MLIP) demonstrates robust predictive capabilities across various physical properties of the system. Figure~\ref{fig:fig_4}c presents the bond angle distributions for sodium-rich ($x = 1.756$) and sodium-poor ($x = 0.02$) $\text{MnFePBA}$ structures at 300~K. The sodium-rich phase exhibits rhombohedral symmetry, characterized by the alternating tilting of $\text{Fe-C}$ and $\text{Mn-N}$ octahedra. Our simulations yield average $\text{Mn-N-C}$ and $\text{Fe-C-N}$ bond angles of 138$^{\circ}$ and 170$^{\circ}$, respectively. These values are in good agreement with previous synchrotron X-ray diffraction (XRD) and neutron powder diffraction measurements, which reported corresponding angles of 141.1$^{\circ}$ and 175$^{\circ}$ \cite{ref2}, and are consistent with DFT calculations by Xiao et al. \cite{ref5}, which yielded Mn–N–C angles of 133°--139° depending on the choice of functional. Further insights into the structural differences are provided by the radial distribution functions (RDFs) of atom pairs in both phases at 300~K, shown in Fig.~\ref{fig:fig_4}d. Distinct differences are observed in the $\text{N-Mn}$, $\text{N-Na}$, and $\text{Fe-Mn}$ interatomic distances. In the sodium-rich structure, the dominant $\text{Na-N}$ Coulombic attraction leads to a shorter average $\text{Na-N}$ bond distance of 2.38~\AA\. This strong interaction concurrently weakens the $d$--$\pi$ orbital overlap between Mn and N atoms, resulting in a longer $\text{Mn-N}$ bond distance (2.17~\AA) consistent with the DFT-reported value of $\sim$2.20 \text{\AA} \cite{Baumgart2026}, and a reduction in electron density within the $\text{C}\equiv\text{N}$ bond \cite{ref3, ref5}. 

Experimental evidence supporting this $\text{C}\equiv\text{N}$ bond weakening comes from \textit{operando} IR-FOEWS measurements by Li \textit{et al.}, where the $\nu_{\text{CN}}$ absorbance band is observed at 2055~cm$^{-1}$ for the sodium-rich phase and 2177~cm$^{-1}$ for the desodiated phase \cite{ref1}. This indicates a weaker $\text{C}\equiv\text{N}$ bond in the sodium-rich state. The prominent $\text{Na-N}$ Coulombic attraction also induces structural distortion, significantly reducing the $\text{Mn-N-C}$ bond angles and, consequently, decreasing the $\text{Mn-Fe}$ interatomic distance.

\begin{figure*}
\includegraphics{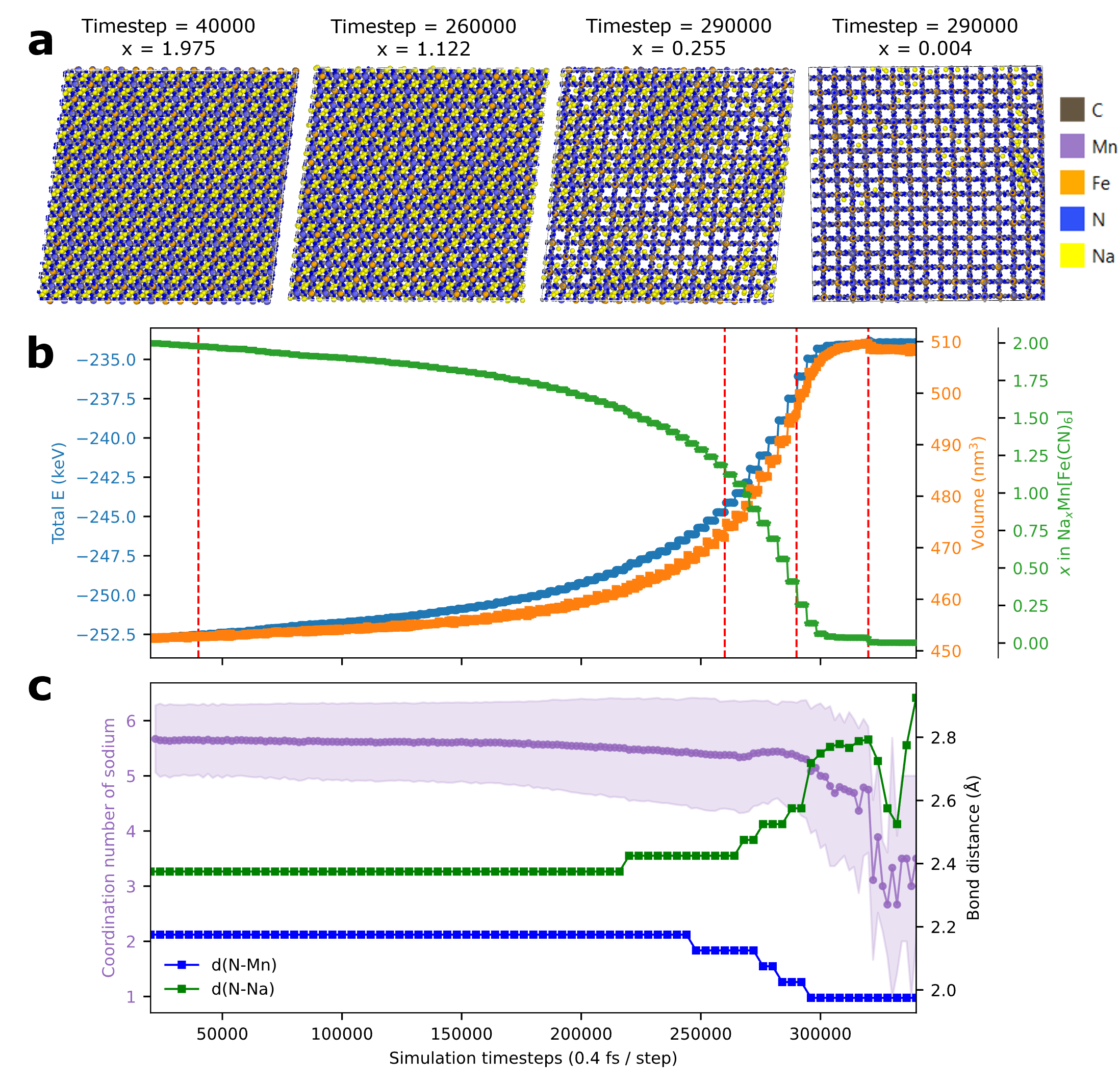}% Here is how to import EPS art
\caption{\label{fig:fig_5}(a) Four snapshots of the simulated system at different sodium concentrations. When fully sodiated, the structure contains 4096 sodium atoms. (b) The evolution of system energy (blue), volume (orange), and sodium concentration (green) profiles during GCMC-MD simulation at 300 K and at µ =-5. During the simulation, for every 4000 NPT timesteps, 2000 sodium insertion/deletion (GC moves) and 2000 sodium translation (MC moves) were performed. The four red dashed lines indicate the time points at which system snapshots were recorded. (c) The evolution of the coordination number of sodium (purple)-—defined as the number of N atoms within a 3.5 Å radius-—and the standard distribution (purple shade), the N-Mn (blue) and N-Na (green) bond distance.}
\end{figure*}

The desodiation process of the $\text{MnFePBA}$ is simulated using GCMC–MD at 300~K at chemical potential, $\mu = -5$. During the simulation, for every 4000 NPT timesteps, 2000 sodium insertion/deletion (GC moves) and 2000 sodium translation (MC moves) were performed. The initial fully sodiated structure containing 4096 sodium atoms exhibits rhombohedral symmetry (Fig.~\ref{fig:fig_4}b). Upon desodiation, the system energy increases, while the structure volume shows a slight expansion and retains rhombohedral symmetry. Initially, sodium atoms have an average coordination number---defined as the number of N atoms within a 3.5~\text{\AA} radius---of 5.7. With progressing desodiation, the average coordination number decreases slightly while the distribution broadens, indicating changes in the local coordination environment of sodium atoms. Simultaneously, the $\text{N--Mn}$ bond length decreases, whereas the $\text{N--Na}$ bond length increases (Fig.~\ref{fig:fig_5}c).

After timesteps 260 000 (130 000 GC and 130 000 MC moves), both system volume and energy rise sharply as desodiation accelerates and the structure undergoes a rhombohedral-to-tetragonal phase transition. During this transition, the $\text{N--Na}$ bond length increases from 2.38~\text{\AA} to 2.9~\text{\AA}, while the $\text{N--Mn}$ bond length decreases from 2.17~\text{\AA} to 1.98~\text{\AA}. As desodiation proceeds further, the average sodium coordination number decreases to three, reflecting a shift in intercalation sites: sodium ions move from the cavity-center sites in the rhombohedral phase to the $48g$ (off-face center, off-FC) and $32f$ (transport hub, TH) \cite{ref35} sites located between face-center ($24d$) and body-center ($8c$) positions in the tetragonal structure (Figure 3b). To validate the MTP's prediction, we performed static DFT geometry optimizations on a $2 \times 2 \times 2$ supercell with a single Na ion and compared the relative stability of a sodium ion at three distinct sites: the TH, off-FC, and FC. The converged energies show that TH is the most stable site for Na in the tetragonal phase, while the off-FC and FC sites are metastable, with energy penalties of 185 and 194~meV, respectively. The trained MTP also predicts the TH site to be the most stable, with energy penalties of 103 and 259~meV for the off-FC and FC sites, respectively.

Notably, this intercalation site differs from those reported in other DFT studies on PBAs, which typically identify face-center sites as the preferred positions \cite{ref3, ref5}. This discrepancy is further contextualized by static DFT calculations on the half-sodiated cubic phase (NaMn[Fe(CN)$_6$]), in which the face-center position is indeed the most stable site. The transport hub position represent higher-energy local minima, with energy penalties of +39~meV/cell \cite{Ito2026}. Similarly, a separate study \cite{Li2025} examining intermediate sodiation levels ($\sim$50\%) found that sodium ions settle into off-center $48g$ positions, driven by a combination of structural and electrostatic factors. At 12.5\% sodiation, the electrostatic environment drives the sodium ion even further off-center toward the reduced transition metal ions. Finally, the stabilization of sodium at the $48g$ and $32f$ sites observed in the present work likely reflects also the dynamic nature of the molecular dynamics simulations, wherein thermal motion at finite temperature and the inherent mobility of sodium ions allow the system to sample the free-energy basins of different sites and favor these off-center positions.

During desodiation, a structural phase transition from rhombohedral to tetragonal symmetry occurs, accompanied by a 12\% volume expansion. This calculated volume expansion aligns well with the 13.9\% expansion reported by previous synchrotron XRD and neutron powder diffraction measurements \cite{ref2}. Beyond structural properties, the MLIP accurately predicts the electrochemical performance of the $\text{MnFe-PBA}$. The average insertion voltage per sodium atom, $V_{\text{ave}}$, is computed from the total energy change due to sodium insertion. It is given by:

\begin{equation}
    V_{\text{ave}} = - \frac{E_f - E_i - n_{\text{Na}} E_{\text{Na}}}{n_{\text{Na}}}
\end{equation}

where $E_i$ and $E_f$ are the potential energy of the initial and the final structures, $E_{\text{Na}}$ is the reference energy of metallic sodium computed by DFT, and $n_{\text{Na}}$ is the number of inserted sodium atoms. The average voltage during desodiation is 3.34~V, which shows excellent agreement with experimental values reported by Song et al., who reported desodiation and sodiation voltages of 3.44~V and 3.53~V, respectively \cite{ref2}.

%%%%%%%%%%%%%%%%%%%%%%%%%%%%%%%%%%%%%%%%%%%%%%%
\subsection{\label{sec:level2}Diffusion properties}

\begin{figure*}[phtb!]
\includegraphics[width=14cm, trim=2 2 2 2, clip]{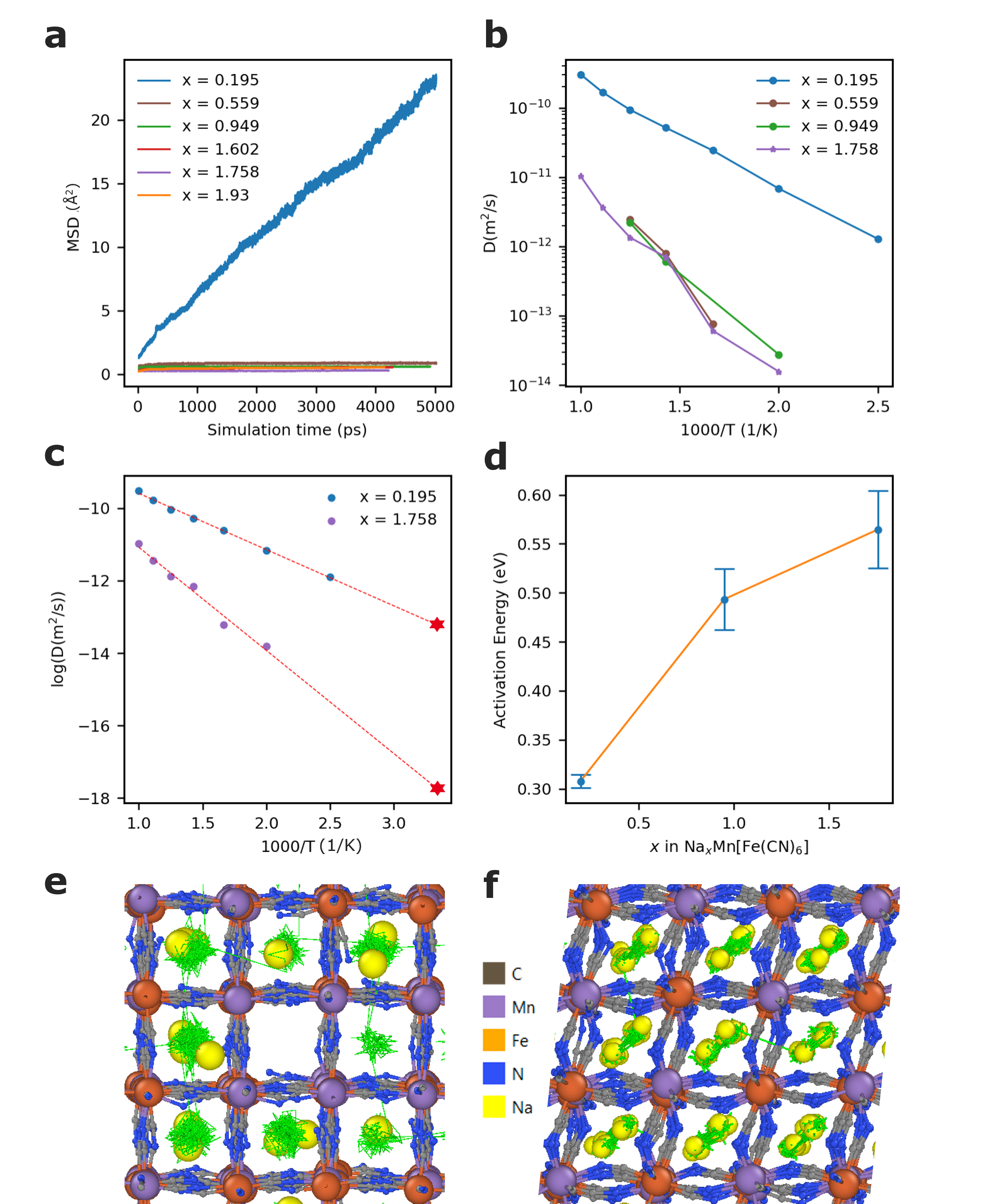}% Here is how to import EPS art
\caption{\label{fig:fig_6}(a) The sodium MSD profiles at 500 K at different sodium concentrations. (b) Arrhenius plot of Na$^+$ diffusivity as a function of 1/T of structures at different sodium concentrations. (c) The fitting of the Arrhenius plot of structures at $x = 0.0195$ and $1.758$. The red dashed lines show the extrapolation of the curve to 300~K (red star markers). The Na$^+$ diffusivity at $x = 0.0195$ and $x = 1.758$ is $7.75 \times 10^{-14}$ and $3.99 \times 10^{-18}$~\si{\meter\squared\per\second}, respectively.(d) The activation energy profile. The error bar is the fitting uncertainty. (e) (f) The Na$^+$ MSD trajectories (colored green) in the sodium-poor and rich structure at 500 K for 5 ns NPT MD simulation. In the tetragonal structure, sodium primarily hops between the 48g or 32f sites within the same MnFePBA cages.}
\end{figure*}

The trained $\text{MLIP}$ was subsequently utilized to compute sodium diffusivity across a range of concentrations and temperatures. The specific configurations for which mean-square displacement ($\text{MSD}$) was calculated are shown in Fig.~S1. Notably, among all investigated structures, only the configuration at $x = 0.195$ exists in the tetragonal phase, while all other structures adopt the rhombohedral phase. Sodium $\text{MSD}$ profiles at varying concentrations reveal striking phase-dependent behavior, as illustrated in Fig.~\ref{fig:fig_6}a. The tetragonal structure at $x = 0.195$ exhibits significantly superior $\text{MSD}$ compared to all other configurations, directly translating to much higher sodium diffusivity. In contrast, the rhombohedral structures demonstrate comparable $\text{MSD}$ profiles and diffusivity regardless of sodium concentration (see Fig.~\ref{fig:fig_6}b). This observation strongly suggests that sodium ion transport is primarily governed by crystal structure rather than concentration within a given phase, with the tetragonal phase facilitating more rapid ionic mobility. Arrhenius fitting of the diffusivity data provides quantitative insight into the activation barriers for sodium migration, shown in Fig.~\ref{fig:fig_6}c. The slope of each fitting line corresponds to the activation energy $E_{\text{a}}$ for sodium diffusion in that particular structure. Extrapolating these fitted functions to 300~K yields predicted $\text{Na}^+$ diffusivities of $7.75 \times 10^{-14}$~m$^2$s$^{-1}$ for the tetragonal phase ($x = 0.195$) and $3.99 \times 10^{-18}$~m$^2$s$^{-1}$ for the rhombohedral phase ($x = 1.758$) -- a difference of approximately four orders of magnitude that underscores the dramatic impact of crystal structure on transport properties. The concentration dependence of activation energy further illuminates these trends, as seen in Fig.~\ref{fig:fig_6}d. The sodium-poor tetragonal phase ($x = 0.195$) exhibits a remarkably low activation energy of approximately 0.3~eV, consistent with facile sodium diffusion through the open framework structure. 

The sodium diffusion trajectories for both structures are presented in Figs.~\ref{fig:fig_6}e and \ref{fig:fig_6}f. In the tetragonal structure, sodium ions primarily exhibit localized motion, hopping between $48g$ or $32f$ sites within the same $\text{PBA}$ cages. Long-range diffusion occurs when sodium migrates from $48g$ or $32f$ sites through the faces of $\text{PBA}$ cages to the corresponding sites in adjacent cages. In contrast, the rhombohedral phase exhibits sodium ion hopping directly between the cage centers. The probability of successful long-range hopping events is significantly lower in the rhombohedral structure, which is consistent with its substantially reduced sodium diffusivity. Moreover, the jump distance is also more favorable to diffusion in the tetragonal phase. In the rhombohedral phase, sodium ions hop directly between cages with an average distance of approximately 3.8~\AA. In contrast, the tetragonal phase exhibits a longer and more broadly distributed jump distance, approximately 4--5~\AA. Conversely, the sodium-rich rhombohedral phases show significantly higher activation energies, ranging from 0.54 to 0.62~eV. 

The observed difference in sodium diffusivity between the tetragonal and rhombohedral phases can be attributed to distinct local environments and interaction strengths. Previous DFT calculations and Bader charge analysis on $\text{Na}_2\text{FeFe(CN)}_6$ by Wang \textit{et al.} \cite{ref3} revealed that the nitrogen ($\text{N}$) atoms in the sodium-rich rhombohedral phase possess a more negative net charge (approximately $-1.47|e|$) compared to the tetragonal phase (approximately $-1.40|e|$). This implies a stronger Coulombic attraction between $\text{Na}^+$ and $\text{N}^-$ ions in the rhombohedral phase. Furthermore, the $\text{Na-N}$ radial distribution function ($\text{RDF}$) profiles in Fig.~\ref{fig:fig_4}b confirm that the rhombohedral phase has a shorter average $\text{Na-N}$ interatomic distance. These combined effects -- stronger Coulombic attraction and shorter bond distances -- result in more restricted sodium movement and thus lower diffusivity in the rhombohedral phase.

In comparison to other simulation studies, Wang et al. performed CI-NEB analysis with a PBE+$U$ functional ($U = 7.0$~eV and 3.0~eV for high-spin and low-spin Fe, respectively) on anhydrous Fe-based PBAs \cite{ref3}. Rhombohedral Prussian White (Na$_2$[FeFe(CN)$_6$]) yields a barrier of 0.44~eV, while the half-filled cubic phase (Na[FeFe(CN)$_6$]) exhibits a strong dependence on the assumed migration mechanism: single-ion hopping through body-center sites along [010] gives 0.64~eV, whereas a cooperative pathway -- in which Na$^+$ ions migrate collectively between adjacent $24d$ sites -- reduces this to 0.40~eV, underscoring that the assumed migration topology is as consequential as the structural model itself. 

Baumgart et al. carried out NEB calculations on desodiated cubic Prussian Yellow (Fe[Fe(CN)$_6$]) \cite{Baumgart2026}, revealing an analogous pathway selectivity: a ladder-like route between face-centered sites carries a barrier of only 60--80~meV, compared with 200--350~meV for direct traversal of the central void. The same study exposes a pronounced exchange--correlation functional sensitivity in rhombohedral Prussian White, where barriers range from $\sim$250~meV to $\sim$500~meV for an identical structure, a factor-of-two spread that renders cross-study comparisons unreliable in the absence of a consistently applied functional. 

While NEB and CI-NEB methods provide well-defined saddle-point energies, they operate on fixed supercells with preset initial and final images and therefore cannot capture phonon coupling, or finite-temperature entropic contributions; the cooperative pathway additionally requires manual image construction, introducing bias into the assumed migration topology. 

Ito et al. conducted DFT-MD simulations with a PBE+$U$ functional ($U = 5.0$~eV for Fe, 4.0~eV for Mn) on fully sodiated cubic Na$_2$Mn[Fe(CN)$_6$], sampling Na$^+$ MSD on a $2 \times 2 \times 2$ supercell over 5~ps, from which an Arrhenius fit extracts a barrier of 92~meV \cite{Ito2026}. In this case, the small supercell and short trajectory could affect correlated diffusion through periodic boundary conditions and limit the statistical convergence of rare hopping events. In our work, MSD sampling is performed on a $16 \times 16 \times 16$ supercell over 5~ns, three orders of magnitude longer in time and substantially larger in spatial extent, ensuring adequate sampling of rare diffusion events and eliminating finite-size artefacts in correlated transport. Computed Na$^+$ migration barriers across PBA structures and simulation frameworks span nearly an order of magnitude, from 60~meV to 640~meV, reflecting both genuine structural diversity and methodological variance that must be disentangled when comparing studies.

%%%%%%%%%%%%%%%%%%%%%%%%%%%%%%%%%%%%%%%%%%%%%%%
\subsection{\label{sec:level2}Interfacial energy and strain energy}

Interfacial energy is a critical thermodynamic parameter governing intra-particle phase separation in electrode materials. A high interfacial energy imposes a significant energy penalty associated with phase coexistence, which can suppress the phase-separation process and stabilize the solid solution state. Consequently, this penalty causes a shrinking of the miscibility gap compared to the bulk material. Furthermore, interfacial energy fundamentally determines the intra-particle phase morphology. For example, the $\text{FP/LFP}$ biphasic system forms striped phase structures due to elastic relaxation near the particle surface \cite{ref7}. This quantity serves as an essential input parameter for continuum modeling: in phase-field models, interfacial energy is described by the Cahn-Hilliard gradient energy density term, which introduces interfacial tension to model a diffuse phase boundary. Critically, the interfacial structure and the corresponding interfacial energies are orientation-dependent. Therefore, accurately determining the structure of a realistic coherent interface is essential for correctly simulating and predicting the associated energy penalties and phase evolution kinetics.

\begin{figure*}[phtb!]
\includegraphics[height=19cm]{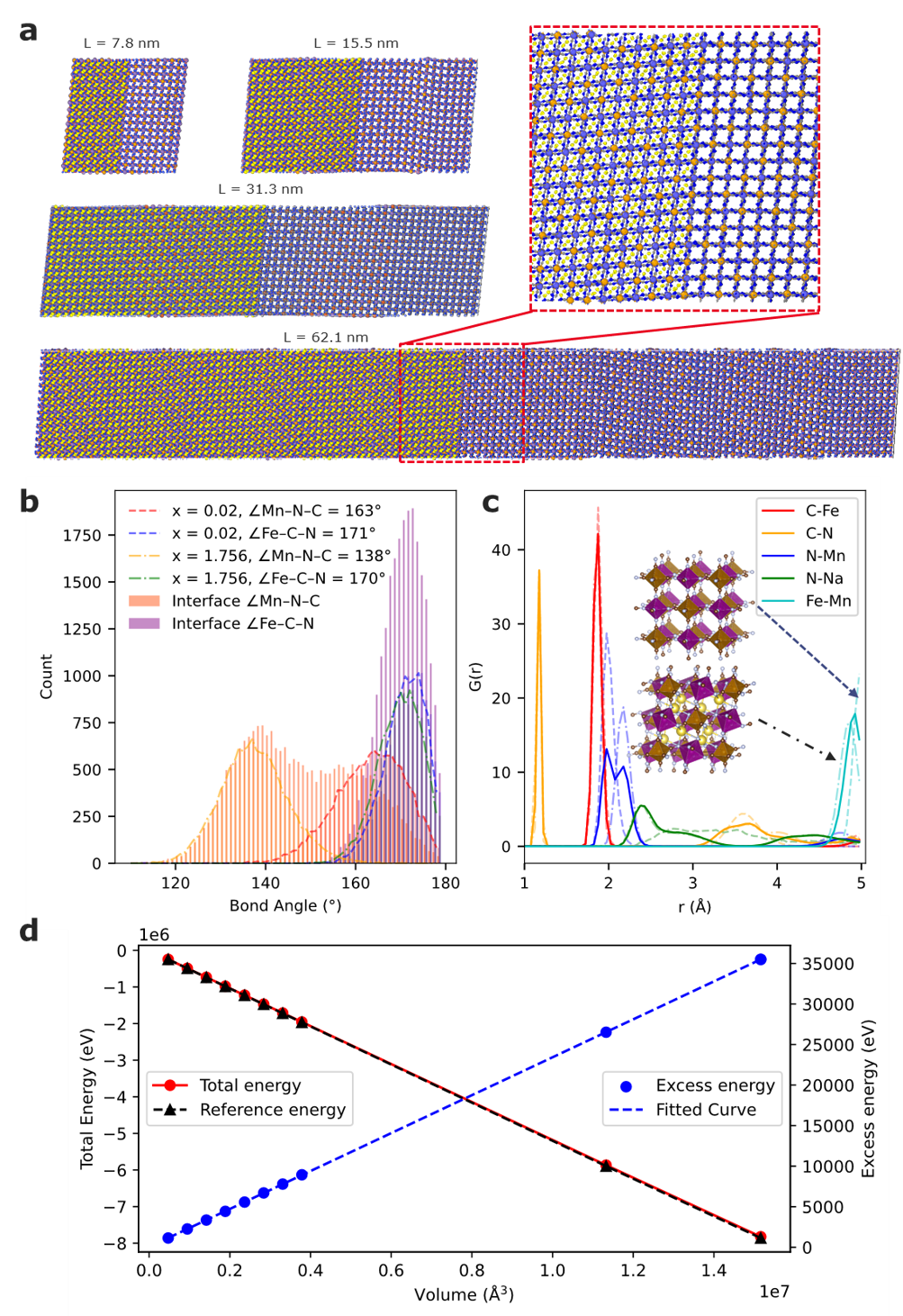}% Here is how to import EPS art
\caption{\label{fig:fig_7}(a) The equilibrated structure of the interfacial structures. (b) The bond angle distribution of the sodium-rich (x = 0.02) and sodium-poor (x = 1.756) structures and the interfacial structure (L = 15.5 nm) (c) Radial distribution functions (RDFs) for various atom pairs in the sodium-poor (dashed lines) and sodium-rich (dash-dot lines) structures and the interfacial structure (solid lines). (d) The total energies (solid red line) and reference energies of each interfacial structure. The difference between the two energies (the excess energy due to having phase separation) is fitted with a linear function (dashed blue line). The interfacial and strain energies are then extracted from the intercept and the slope of the fitted linear function.}
\end{figure*}

The sodium-rich and sodium-poor phases exhibit rhombohedral and tetragonal symmetry, respectively. Although both phases have similar $\text{Fe--Mn}$ distances, their lattice symmetries differ substantially. The rhombohedral phase is characterized by lattice angles $\alpha = \beta = \gamma = 82.6^{\circ}$, whereas the tetragonal phase adopts $\alpha = \beta = \gamma = 90^{\circ}$. Consequently, forming a coherent interface necessitates lattice distortion in both phases to accommodate this symmetry mismatch. Structural analysis reveals the extent of these distortions. In the structure with $L = 7.8$~nm (Fig.~\ref{fig:fig_7}a), the non-sodiated side deviates measurably from tetragonal symmetry. Bond angle analysis of the $L = 15.5$~nm structure (Fig.~\ref{fig:fig_7}b) demonstrates that achieving interfacial coherency distorts both regions: $\text{Mn--N--C}$ bond angles in the sodiated phase are larger than those in the bulk sodiated phase (dash-dotted lines), whereas those in the non-sodiated phase are reduced compared to the bulk non-sodiated phase (dashed lines). The more pronounced deviation in the non-sodiated region indicates greater structural flexibility in this phase. Without the presence of sodium ions to occupy the cavities and stabilize the framework via Coulombic interactions, the PBA cages are more susceptible to the mechanical distortions required to maintain lattice coherency within the rhombohedral phase. Radial distribution function (RDF) analysis (Fig.~\ref{fig:fig_7}c) confirms that the interfacial structure exhibits two distinct $\text{N--Mn}$ and $\text{N--Na}$ bond lengths corresponding to the sodiated and desodiated phases, while the $\text{Fe--Mn}$ distance assumes an intermediate value between the two extremes. For interfacial structures with $L > 7.8$~nm, domains of different crystal orientations emerge to relieve strain associated with the coherent interface. Remarkably, even at $L = 62.1$~nm -- an eightfold larger system —- the non-sodiated phase remains distorted, and no relaxed tetragonal domain is observed. This persistent distortion underscores the long-range nature of strain fields induced by the phase boundary.

Interfacial and strain energies were quantified by evaluating the excess energy associated with phase separation and fitting this excess energy against cell volume (Fig.~\ref{fig:fig_7}d). The interfacial energy of the coherent interface is $2.828 \pm 1.71$~mJ$\cdot$m$^{-2}$, comparable to that of the $ac$ coherent interface in $\text{LFP--FP}$ ($\sim$7~mJ$\cdot$m$^{-2}$) \cite{ref25}. In contrast, the strain energy is significantly larger at $37.655 \pm 0.0875$~MJ$\cdot$m$^{-3}$, compared to $\text{LFP}$ values of 3--15~MJ$\cdot$m$^{-3}$ depending on particle morphology. This relatively low interfacial energy arises from the flexibility of the cyanide groups linking the Mn and Fe redox centers, thereby facilitating the lattice matching between the rhombohedral ($a = b = c = 9.564$~\AA) and tetragonal ($a = b = 9.986, c = 9.983$~\AA) phases despite their differing lattice parameters. Conversely, the substantially higher strain energy originates from the significant difference in lattice symmetry: maintaining a coherent interface induces structural distortion throughout the entire material, particularly in the more deformable, non-sodiated phase.

Our approach offers a general method to construct coherent interfaces without specific knowledge of the material system and careful treatment of boundary conditions. Previous computational studies on $\text{LFP}$ \cite{ref25} and $\text{LCO}$ \cite{ref12} artificially joined two phases together by imposing constraints such as matching lattice parameters along the interfacial plane, then performed structural relaxation with fixed cell vectors. This approach may produce artificial interfaces that differ from those formed under equilibrium conditions. In contrast, our trained $\text{MLIP}$ enables $\text{NPT MD}$ simulations of interfacial systems containing hundreds of thousands of atoms to relax without imposing structural constraints. The resulting coherent interface thus emerges naturally from the energy and strain relaxation of the system, providing a more realistic representation of phase boundaries in $\text{MnFe-PBA}$.

%%%%%%%%%%%%%%%%%%%%%%%%%%%%%%%%%%%%%%%%%%%%%%%
\subsection{\label{sec:level2}Free energy sampling}

At each temperature, biased $\text{GCMC-MD}$ simulations were performed across 82 target sodium concentrations, spaced by 50 Na units and ranging from 200 to 4300. The initial structure was in the tetragonal phase and contained 5 sodium atoms. A harmonic biasing potential with $N_{\text{Na}}^t$ and $\phi = 0.002$ was applied to guide the sodium insertion process. Fig.~\ref{fig:fig_S4}a illustrates the sampling procedure at 400~K, with the inset highlighting $\text{GCMC}$ trajectories at four selected $N_{\text{Na}}^t$ values. At each $N_{\text{Na}}^t$, hybrid $\text{GCMC-MD}$ simulations cyles composed of 4000 $\text{NPT}$ timesteps and 2000 GC moves and 2000 MC moves were run repeatedly until the sodium content converged to an equilibrium value. Upon convergence, the corresponding chemical potential was computed, and the simulation proceeded to the next $N_{\text{Na}}^t$. To ensure that true equilibrium sodium concentrations were reached, each sampling was repeated multiple times (Fig.~\ref{fig:fig_S4}b, c). The evolution of sodium content across these repeated samplings is shown in Fig.~\ref{fig:fig_S4}d, demonstrating that extended sampling -- achieved by successive resampling —- is essential for capturing the true equilibrium concentration. Fig.~\ref{fig:fig_S4}e, presents the chemical potential profiles from each sampling iteration, clearly showing that longer sampling significantly reduces noise and mitigates overestimation of the chemical potential.

\begin{figure*}
\includegraphics{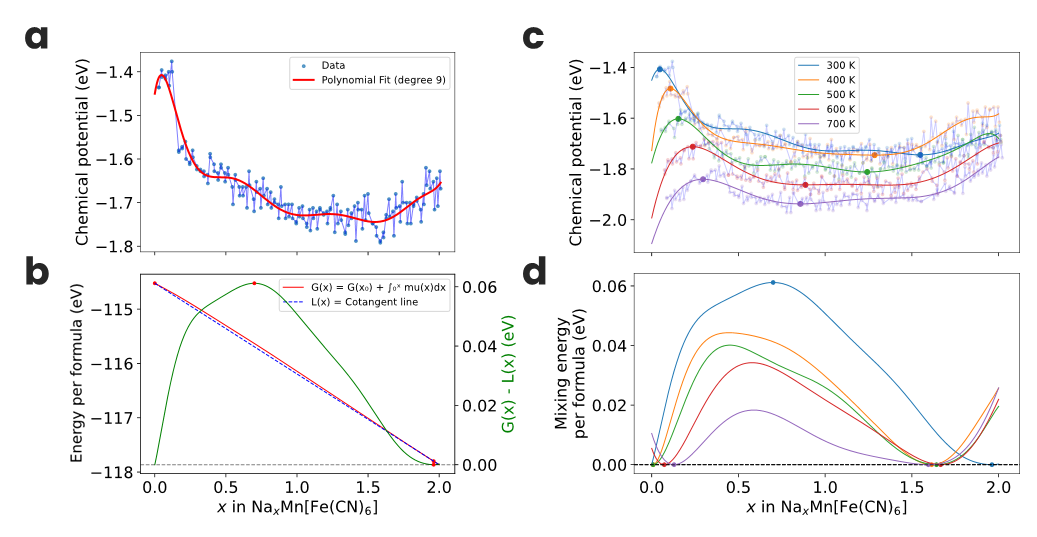}% Here is how to import EPS art
\caption{\label{fig:fig_8}(a) Chemical potential profiles for sodium as a function of composition, sampled at 300 K (blue markers) with polynomial fits (red lines) used for thermodynamic integration. (b) Free energy profiles (red) obtained by integrating the fitted chemical potential curves. Dashed blue lines indicate the common tangent construction identifying coexisting equilibrium phases. Green curves represent the mixing energy, illustrating the energetic penalty of solid solution formation. (c) Chemical potential profiles at different temperatures, showing sampled data (markers) and polynomial fits (solid lines). The dots mark the spinodal points, which bound the spinodal region where phase decomposition occurs spontaneously. (d) Mixing energy profiles at different temperatures. Dots indicate binodal points that define the boundaries of the miscibility gap.}
\end{figure*}

The sampled chemical potential profiles are fitted using polynomial functions (Fig.~\ref{fig:fig_8}a). These fitted functions are then integrated with respect to sodium content to obtain the free energy profiles shown in Fig.~\ref{fig:fig_8}b. Convex hull analysis of the free energy profiles yields the common tangent construction (dashed blue lines), which connects the equilibrium free energies of the two coexisting phases and identifies the binodal points that define the miscibility gap. The mixing free energy (green curves) represents the energetic penalty of forming a homogeneous solid solution relative to phase-separated states. Sampled and fitted chemical potential profiles spanning 300~K to 700~K are compiled in Fig.~\ref{fig:fig_8}c. Spinodal points, identified where $\partial\mu/\partial x = 0$, mark the boundaries of the spinodal region within which phase decomposition occurs spontaneously. The chemical spinodal boundary exhibits pronounced asymmetry: at 300~K, spinodal points occur at $x = 0$ and $x = 1.95$, and with increasing temperature, the upper boundary shifts to lower sodium concentrations far more substantially than the lower boundary shifts to higher concentrations.

Mixing energy profiles for homogeneous solid solution formation are shown in Fig.~\ref{fig:fig_8}d. At 300~K and $x = 0.7$, the mixing energy reaches 0.06~eV per formula unit -— significantly higher than the 0.01~eV per formula unit reported for the $\text{LFP-FP}$ system \cite{ref7}. The binodal points at 300~K, determined from common tangent construction, occur at $x = 0$ and $x = 1.95$, yielding a wide miscibility gap consistent with the single-plateau feature observed in the material's open-circuit voltage ($\text{OCV}$) curve \cite{ref1, ref2}.

The asymmetric mixing energy profile, characterized by a gentler slope at high sodium concentrations, indicates that compositional fluctuations are more energetically favorable in the sodiated state. This prediction aligns with experimental rate capability measurements showing that at high C-rates, overpotential increases and voltage plateau diminution occur more dramatically on the high-concentration side \cite{ref2} -— a phenomenon directly attributable to the free energy characteristics revealed by our calculations. With increasing temperature, both mixing energy and miscibility gap decrease gradually, consistent with expected behavior for spinodal systems.

%%%%%%%%%%%%%%%%%%%%%%%%%%%%%%%%%%%%%%%%%%%%%%%
\subsection{\label{sec:level2}P2D phase field simulation}
\begin{figure*}[phtb!]
\includegraphics[height=19cm]{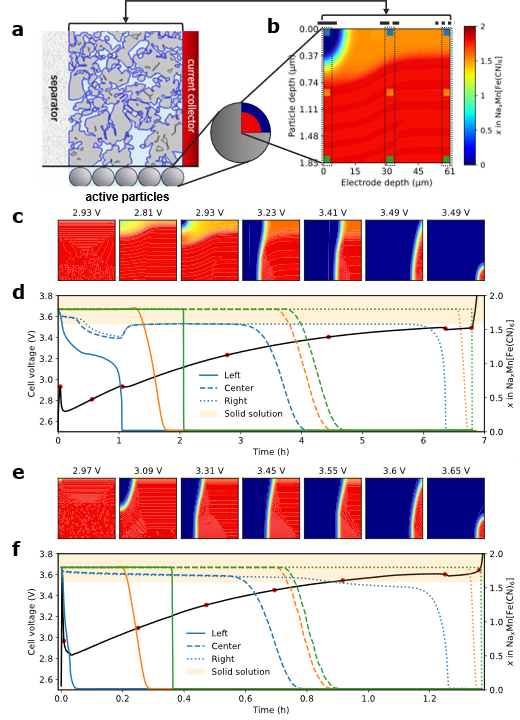}% Here is how to import EPS art
\caption{\label{fig:fig_9}(a) Schematic of the pseudo-two-dimensional (p2D) phase-field system. (b) Sodium concentration map obtained from the p2D phase-field simulation, where the x-axis denotes the electrode depth (distance from the separator) and the y-axis denotes the particle depth (distance from the spherical particle surface toward the center). Nine representative regions are highlighted (blue, green, and orange), spanning from the particle surface to the particle center; the solid, dashed, and dotted vertical lines indicate positions from the separator to the current collector, respectively. (c,e) Sodium concentration maps at charging rates of 0.1C and 0.5C, respectively. (d,f) Corresponding simulated cell voltages (solid black lines), with red markers indicating the voltages at which the concentration maps in (c) and (e) were extracted. The blue, green, and orange curves (solid, dashed, and dotted lines) show the temporal evolution of sodium concentration in the nine marked regions defined in (b). The orange-shaded region denotes the solid-solution concentration range determined from the binodal points of the free-energy profile at 400 K (see Figure 8d).}
\end{figure*}

Pseudo-2D simulations with the phase-field model at the positive electrode are performed on a cell with MnFePBA active material at the positive electrode and hard carbon at the negative electrode, using the software COMSOL Multiphysics\textsuperscript{\textregistered}. The positive electrode thickness is 61~\si{\micro\meter} with a porosity of 25\%, and the negative electrode thickness is 100~\si{\micro\meter} with a porosity of 28\%. The MnFePBA active particle radius is 1.85~\si{\micro\meter} (Fig.~\ref{fig:fig_9}b), and the hard carbon particle radius is 8~\si{\micro\meter}. The cell design parameters used in the simulation do not correspond to any commercial cell design but can be considered realistic values to emphasize the main phenomena at the electrode scale stemming from the phase-field dynamics. The simulation corresponds to a full charge from 2.5~V to 4.5~V, with the MnFePBA material initially exhibiting a homogeneous sodium concentration at $x = 1.8$, above the concentration of the binodal point. The initial stoichiometry of hard carbon is $x = 0.05$.

All the bulk kinetic properties of the PBA material -- that is, the chemical potential, the diffusivity, and the interface energy density -- are deduced from the Monte Carlo simulations. The only missing kinetic parameter needed to obtain a full upscaled kinetic model of PBA is the exchange current density of the Butler-Volmer law. This parameter has been set to a realistic value of $j_0 = 1$~\si{\ampere\per\meter\squared}. As explained in Section~\ref{section:interface_energy_computation}, the real interface energy value close to 5~\si{\milli\joule\per\meter\squared} cannot be implemented in the simulation because it would lead to a vanishingly thin interface compared to the particle radius. A value of 500~\si{\milli\joule\per\meter\squared} was chosen in the simulation to maintain a typical interface width of one-tenth of the particle radius. The artificial increase in the interface energy has no macroscopic consequences on the simulation results because the total interface energy in the porous electrode is far smaller than the electrochemical energy exchanged during charging. Considering the most unfavorable case where all particles have nucleated, the maximal interface energy per unit area of electrode in the simulation is 35~\si{\joule\per\meter\squared}. By contrast, the electrochemical energy stored in the PBA material is on the order of 400~\si{\kilo\joule\per\meter\squared}, four orders of magnitude larger.

Fig.~\ref{fig:fig_9} displays the simulation results of charging at two C-rates, 0.1C and 0.5C, at a temperature of 400~K. A detailed description of the phase-field dynamics at the positive electrode is provided in the following.

Upon charging at 0.1C, the cell voltage initially surged from 2.5~V to 2.93~V (black curve in Fig.~\ref{fig:fig_9}d). This surge is attributed both to the energy barrier of initiating nucleation of the sodium-poor phase and to the very strong kinetic limitation of sodium transport in the sodium-rich phase. The solid diffusion coefficients are $1.9\times10^{-18}$~\si{\meter\squared\per\second} in the sodium-rich phase and $6.13\times10^{-14}$~\si{\meter\squared\per\second} in the sodium-poor phase, respectively. Desodiation subsequently commences as a typical solid-solution reaction at the particle surface throughout the electrode. Particles located near the separator (the upper left boundary in the P2D domain) desodiate more rapidly due to being more accessible to the electrolyte, which is why nucleation first starts at the separator boundary. Once a region of sodium-poor phase has been created, sodium can be removed more efficiently from the sodium-rich phase by flux through the sodium-poor phase, and the overpotential drops.

At 2.93~V, the cell voltage plateaus as the sodium concentration at the particle surface (blue solid line) rapidly drops to the binodal concentration, $x = 0.02$ (Fig.~\ref{fig:fig_9}c and d). This drop signals the onset of the biphasic reaction and the formation of the biphasic front. The particle surfaces deeper in the electrode that originally underwent a solid-solution reaction quickly recover their concentration back to the binodal concentration, $x = 1.58$ (blue dashed and blue dotted lines), thereby mitigating the energy penalty associated with phase mixing. Contrary to particle-scale LFP simulations, in which the miscibility gap shrinks to avoid the energy penalty resulting from forming an interface \cite{ref7}, the MnFePBA miscibility gap does not shrink in the P2D domain, suggesting that strain influence on equilibrium boundaries is less significant at the electrode scale.

As nucleation starts, desodiation predominantly occurs through the biphasic reaction, with only a tiny fraction of the particle surface undergoing a solid-solution reaction to supply the necessary sodium flux. This preference for the biphasic mechanism is both kinetically and energetically favorable, particularly because the sodium-rich phase has an extremely low sodium diffusivity of $1.9\times10^{-18}$~\si{\meter\squared\per\second} compared to the significantly higher sodium diffusivity of $6.13\times10^{-14}$~\si{\meter\squared\per\second} in the sodium-poor phase. Given this kinetic hindrance, the two-phase reaction path is the dominant mode for achieving the required high sodium flux.

As charging proceeds, the cell voltage gradually rises due to the OCV of the hard carbon material, and the biphasic front propagates both across the electrode thickness and towards the core of the active particle. The desodiation process concludes when the cell voltage reaches 3.49~V, at which point all particle surfaces are fully desodiated, removing the interparticle interface and resulting in a drop in the cell voltage. The simulation highlights that the main desodiation pathway for the PBA electrode is the particle-by-particle dynamics, that consists in primarily fully desodiating the particles near the separator before initiating nucleation gradually deeper into the electrode. The main concentration heterogeneity is directed along the electrode depth, whereas the sodium concentration profile inside particles at a given electrode depth is rather homogeneous.

At a higher rate of 0.5C, the charging process exhibits similar phenomenology, yet with some kinetic differences. Upon charging, the cell voltage surges to a significantly higher potential of 3.66~V (Fig.~\ref{fig:fig_9}e). The particle surfaces undergo a much shorter solid-solution reaction compared to 0.1C. The nucleation of the sodium-poor phase starts quickly, forming a biphasic front at 2.85~V rather than 2.93~V as observed at 0.1C. This indicates that the initial voltage surge is sufficient to overcome the nucleation energy barrier more rapidly and initiate the phase transition. At this higher rate, the particle surface concentration (dashed and dotted blue lines) plateaus at $x = 1.68$, while the sodium-poor phase plateaus at $x = 0.017$. Contrary to the 0.1C kinetics, the particles deep in the electrode do not have time to reach the border of the solid-solution phase before the first particle at the electrode surface nucleates. They thus remain close to the initial concentration $x = 1.8$ until the phase front reaches them. The concentration value in the sodium-poor phase deviates from the equilibrium binodal concentration ($x = 0.02$). This deviation is attributed to the higher overpotential at 0.5C, where the cell voltage is consistently approximately 1.1~V higher than that at 0.1C. This elevated overpotential stabilizes the non-equilibrium concentration at $x = 0.017$. The charge ends at 3.6~V, where all particle surfaces are fully desodiated, similar to the 0.1C charging process. The simulation shows that the desodiation mechanism at 0.5C is similar to that at 0.1C. It is still kinetically favorable to fully desodiate particles near the separator before propagating nucleation deeper into the electrode. The concentration heterogeneity is still directed along the electrode depth.

The particle-by-particle desodiation mechanism typically happens when the potential barrier to overcome to start nucleation – that is the potential difference between the binodal and the spinodal points – is large compared to the overpotential due to sodium transport. Such phenomenology will be observed also at higher temperatures because the nucleation potential barrier decreases slower with temperature than the overpotential due to sodium transport. The main modification when raising the temperature is that the spinodal point of the sodium-rich phase is shifted from x$\approx$1.3 at 400K to x$\approx$0.85 at 700K. The duration of the metastable phase at beginning of charge will thus last longer at 700K than at 400K.  From a practical point of view, it should be borne in mind that PBA decomposes at 500K under N or Ar \cite{ref2}. A simulation at temperatures higher than 500K cannot represent a realistic electrode operation.

As it stands, the pseudo-2D model cannot represent a realistic discharge of the PBA material. The P2D model is built on a core-shell representation of the sodium concentration in active particles. Since the sodium-rich phase has a significantly low sodium diffusivity, once this phase nucleates and covers the particle surfaces, the low diffusivity effectively prevents further desodiation during the simulation. It can be intuited that the real sodiation mechanism in a 3-dimensional particle still occurs through the sodium-poor phase, which implies that both phases coexist at the particle surface until the particle is fully sodiated. More elaborate models can be used to release the isotropic constraint. A P3D model can be built with a 3-dimensional solid domain representing the electrode depth coupled with 2-dimensional particles. Another approach would be to directly simulate the phase-field dynamics in a microstructure model, including the pores geometry. However, the numerical cost of such models is much larger.

\section{\label{sec:level1}Conclusion}
In this work, we successfully developed a generalizable scale-bridging computational framework that enables predictive modeling of insertion-type electrode materials, specifically MnFePBA, from atomic to electrode scales without reliance on empirical inputs. The key innovation lies in the methodology: an active-learning strategy that trains a Moment Tensor Potential (MTP)-based machine-learning interatomic potential (MLIP) through iterative GCMC-MD sampling. This approach robustly captures diverse configuration spaces, overcoming the fundamental limitations of rigid-lattice models such as cluster expansion methods.

Using the validated MTP-MLIP, we accurately predicted essential physical properties that are critical for understanding electrochemical performance. We determined the concentration- and phase-dependent diffusivities for both the sodium-rich (rhombohedral) and sodium-poor (tetragonal) phases, quantifying the drastic kinetic difference between them. In the sodium poor phase, sodium averagely coordinates with four instead of six nitrogen neighbors and has a longer Na-N interatomic distance and a weaker Coulombic attraction between Na and N, compared with the sodium rich phase, allowing for a higher mobility. As a result, the sodium diffusivity varies by four orders of magnitude between phases: $7.75 \times 10^{-14}$~\si{\meter\squared\per\second} in the sodium-poor phase compared to $3.99 \times 10^{-18}$~\si{\meter\squared\per\second} in the sodium-rich phase at 300~K.

Furthermore, we quantified the coherent interfacial and strain energies associated with phase separation, finding that MnFePBA exhibits lower interfacial energy but higher strain energy than the LiFePO$_4$ (LFP) system. This behavior results from the flexible lattice framework composed of cyanide groups and the significant difference in lattice symmetry between the rhombohedral and tetragonal phases. We also reconstructed the complete free-energy landscape via biased GCMC-MD sampling, successfully predicting the double-well feature characteristic of a biphasic system.

These values obtained from atomistic insights were then directly incorporated into a pseudo-2D (P2D) phase-field simulation to model phase boundary propagation and rate-dependent phenomena across the electrode. The P2D simulations revealed that desodiation proceeds through a particle-by-particle mechanism, with particles near the separator fully desodiated before nucleation propagates deeper into the electrode. The concentration heterogeneity is primarily directed along the electrode depth rather than within individual particles. The particle isotropy constraint in the P2D model prevents any realistic simulation of the sodiation process, due to the extremely low diffusivity of the sodium-rich phase. A continuation of the present work is ongoing to include the complex phase-field dynamics of PBA in a pseudo-microstructure model that would go beyond the intrinsic limitations of the P2D.

This framework successfully links atomistic energetics and kinetics to continuum-scale behavior, demonstrating its capability to build a model at the device level from first-principles. By elucidating the thermodynamic drivers (e.g., high mixing energy and strain energy) and kinetic limitations (e.g., extremely low sodium-rich phase diffusivity), this work not only offers critical insights into the performances of MnFePBA but also establishes a blueprint for the accelerated computational design and optimization of next-generation insertion-type electrode materials. The methodology is fully generalizable and can be applied to other phase-separating electrode materials, enabling rational materials design guided by first-principles predictions.
\clearpage 
%bibliography{apssamp}% Produces the bibliography via BibTeX.


% In your document:
\clearpage 
\appendix
\section{Computation of the interface energy in the phase-field model}

\renewcommand{\thefigure}{A\arabic{figure}}
\renewcommand{\thetable}{A\arabic{table}}
\renewcommand{\theequation}{A\arabic{equation}}
\setcounter{figure}{0}
\setcounter{table}{0}
\setcounter{equation}{0}

\label{section:interface_energy_computation}
The interface energy can be computed from the excess free energy in the presence of a planar interface, as defined by equation \ref{eq:eq_25}
\begin{eqnarray}
\Delta \mathcal{G} && = \int_{V} \frac{k}{2} |\nabla N^{eq}|^2 + G(N^{eq}) \, d^3r \nonumber \\
&& - \int_{V_1} G(N^1) \, d^3r - \int_{V_2} G(N^2) \, d^3r
\end{eqnarray}

with the values of the volumes $V_1$ and $V_2$ given by the relation of mass conservation
\begin{equation}
\int_{V} N^{eq} \, d^3r = N^1 V_1 + N^2 V_2
\label{eq:eq_A2}\end{equation}

The subscript $N_a$ is omitted for clarity.
The equilibrium concentration profile $N^{eq}$ in an infinite domain is the solution in one dimension of the stationary problem
\begin{equation}
\nabla \frac{\delta \mathcal{G}}{\delta N_a} = 0
\end{equation}
which is equivalent to

\begin{equation}
-k \frac{d^2 N^{eq}}{dr^2} + \mu(N^{eq}) = \mu(N^1)
\end{equation}

The chemical potential satisfies $\mu(N^1) = \mu(N^2)$ because both phases are at equilibrium. The integration of the previous equation can be done by multiplying both sides by $\frac{dN^{eq}}{dr}$ and using the asymptotic condition that $N \to N^1$ on one side away from the interface. It gives

\begin{equation}
-\frac{k}{2} \left| \frac{dN^{eq}}{dr} \right|^2 + G(N^{eq}) = \mu(N^1) \cdot (N^{eq} - N^1) + G(N^1)
\end{equation}

We then introduce the function defined by equation \ref{eq:eq_27}

\begin{equation}
p(N) = G(N) - G(N^1) - \mu(N^1) \cdot (N - N^1)
\end{equation}

\begin{figure*}
\includegraphics{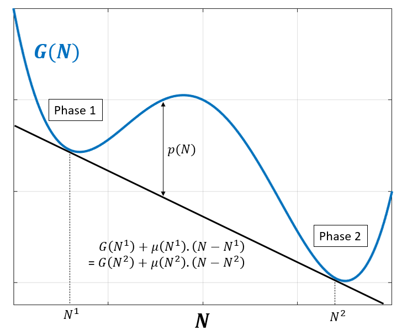}% Here is how to import EPS art
\caption{\label{fig:fig_S5}Shematic representaion of the non-convex part of the Gibbs free energy (blue curve). The common tangent construction gives the equilibrium concentrations N$^1$ and N$^2$ of the two phases (black line).}
\end{figure*}

The common tangent construction of the free energy (see Figure \ref{fig:fig_S5}) shows that the function $p$ can be equivalently written

\begin{equation}
p(N) = G(N) - G(N^2) - \mu(N^2) \cdot (N - N^2)
\end{equation}

We thus obtain the stationary concentration profile as the solution of

\begin{equation}
\frac{k}{2} \left| \frac{dN^{eq}}{dr} \right|^2 = p(N^{eq})
\label{eq:eq_A8}\end{equation}

The excess of free energy can then be written

\begin{eqnarray}
\Delta \mathcal{G} && = \int_{V} \frac{k}{2} |\nabla N^{eq}|^2 d^3r \nonumber\\ 
&& + \int_{V_1} [G(N^{eq}) - G(N^1)] d^3r\nonumber\\ 
&& + \int_{V_2} [G(N^{eq}) - G(N^2)] d^3r
\end{eqnarray}

Using the expression of $p(N)$, we get

\begin{eqnarray}
\Delta \mathcal{G} && = \int_{V} \frac{k}{2} |\nabla N^{eq}|^2 d^3r \nonumber\\
&& + \int_{V_1} [p(N^{eq}) + \mu(N^1) \cdot (N^{eq} - N^1)] d^3r \nonumber\\
&& +\int_{V_2} [p(N^{eq}) + \mu(N^2) \cdot (N^{eq} - N^2)] d^3r
\end{eqnarray}

\begin{eqnarray}
\Delta \mathcal{G} && = 
\int_{V} \left( \frac{k}{2} |\nabla N^{eq}|^2 +p(N^{eq}) \right) d^3r \nonumber\\ &&
+ \mu(N^1) \int_{V_1} (N^{eq} - N^1) d^3r \nonumber\\ &&
+ \mu(N^2) \int_{V_2} (N^{eq} - N^2) d^3r
\end{eqnarray}

With the equality of the chemical potentials $\mu(N^1) = \mu(N^2)$, we can group the last two terms. The function $p$ can again be replaced using equation \ref{eq:eq_A8}

\begin{eqnarray}
\Delta \mathcal{G} && = \int_{V} k |\nabla N^{eq}|^2 d^3r\nonumber\\
&& + \mu(N^1) \left[ \int_{V} N^{eq} d^3r - N^1 V_1 - N^2 V_2 \right]
\end{eqnarray}

The last term vanishes due to the relation of mass conservation given by equation \ref{eq:eq_A2}. We finally express the interface energy density $\gamma$ by considering the excess free energy for a given area $S$ in the interface plane

\begin{equation}
\gamma = \Delta \mathcal{G} / S = \int_{-\infty}^{+\infty} k \left| \frac{dN^{eq}}{dr} \right|^2 dr
\end{equation}

The final step is to do the change of variable $N = N^{eq}(r)$, and replace $\frac{dN^{eq}}{dr}$ with the relation \ref{eq:eq_A8}

\begin{equation}
\gamma = \int_{N^1}^{N^2} k \sqrt{\frac{2p(N)}{k}} dN
\end{equation}

which corresponds to equation \ref{eq:eq_26}.

\section{MLFF-AIMD and MLIP-MD}

Additional figures related to the construction of the force field and simulations are presented below.

\renewcommand{\thefigure}{B\arabic{figure}}
\renewcommand{\thetable}{B\arabic{table}}
\renewcommand{\theequation}{B\arabic{equation}}
\setcounter{figure}{0}
\setcounter{table}{0}
\setcounter{equation}{0}

\begin{figure*}
\includegraphics{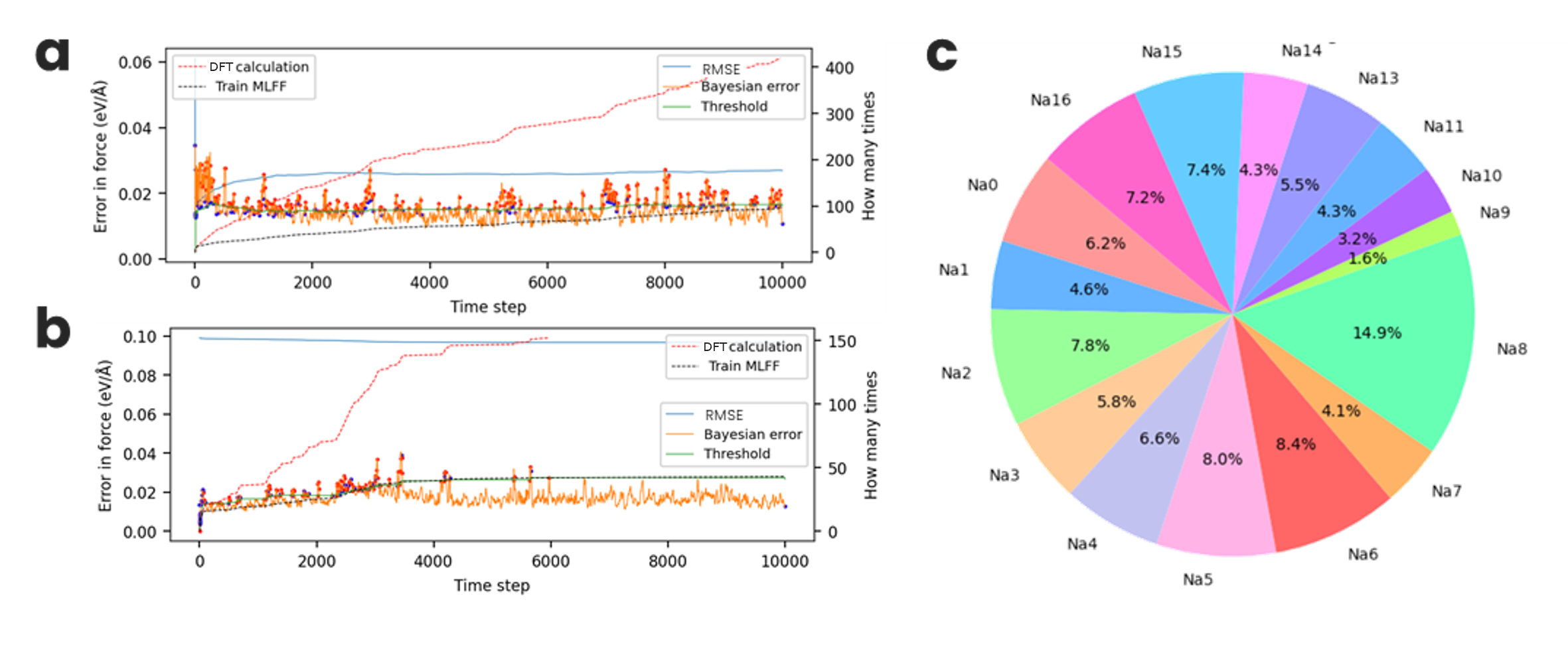}% Here is how to import EPS art
\caption{\label{fig:fig_S1}The MLFF-AIMD Bayesian error (uncertainty) profile (orange) during (a) the first and (b) the second NPT-MD simulations. The red dots mark when the uncertainty of the structure is greater than the threshold (green). The black dots show when the on-the-fly training is performed. The number of DFT calculations and the number of trainings performed during NPT-MD are shown in dashed red and black. (c) The pie chart shows the data distribution of the pre-training set generated by MLFF-AIMD}
\end{figure*}

\begin{figure*}
\includegraphics{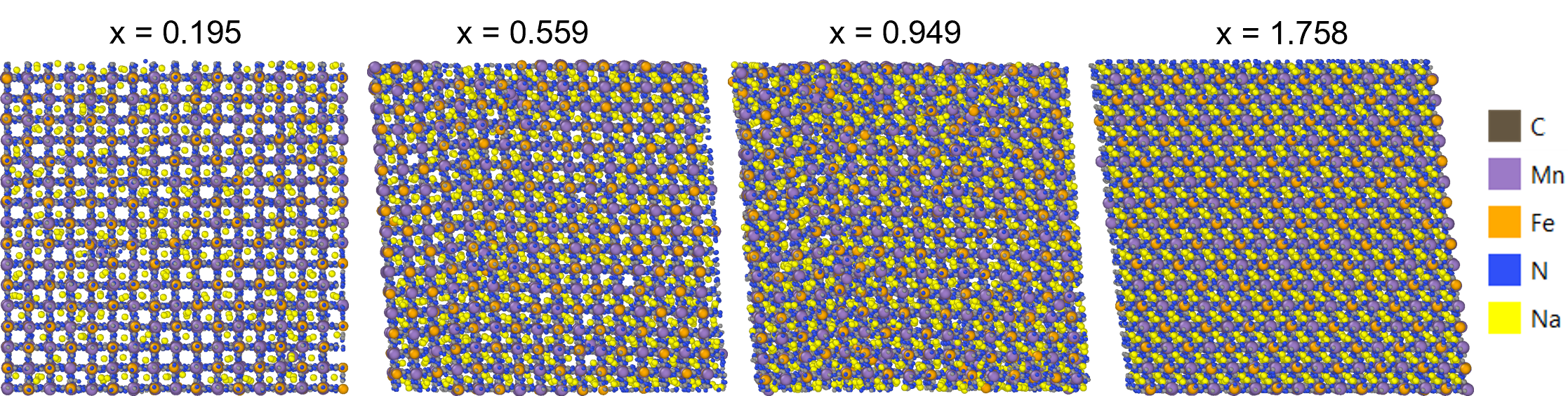}% Here is how to import EPS art
\caption{\label{fig:fig_S2}Snapshots of equilibrated structures at different sodium concentrations prepared by GCMC (de)sodiation. NPT simulations were subsequently performed on these configurations to extract the sodium mean-square displacement (MSD) for diffusivity calculations. Only the structure at x = 0.195 exhibits tetragonal symmetry.}
\end{figure*}

\begin{figure*}[phtb!]
\includegraphics[height=14cm]{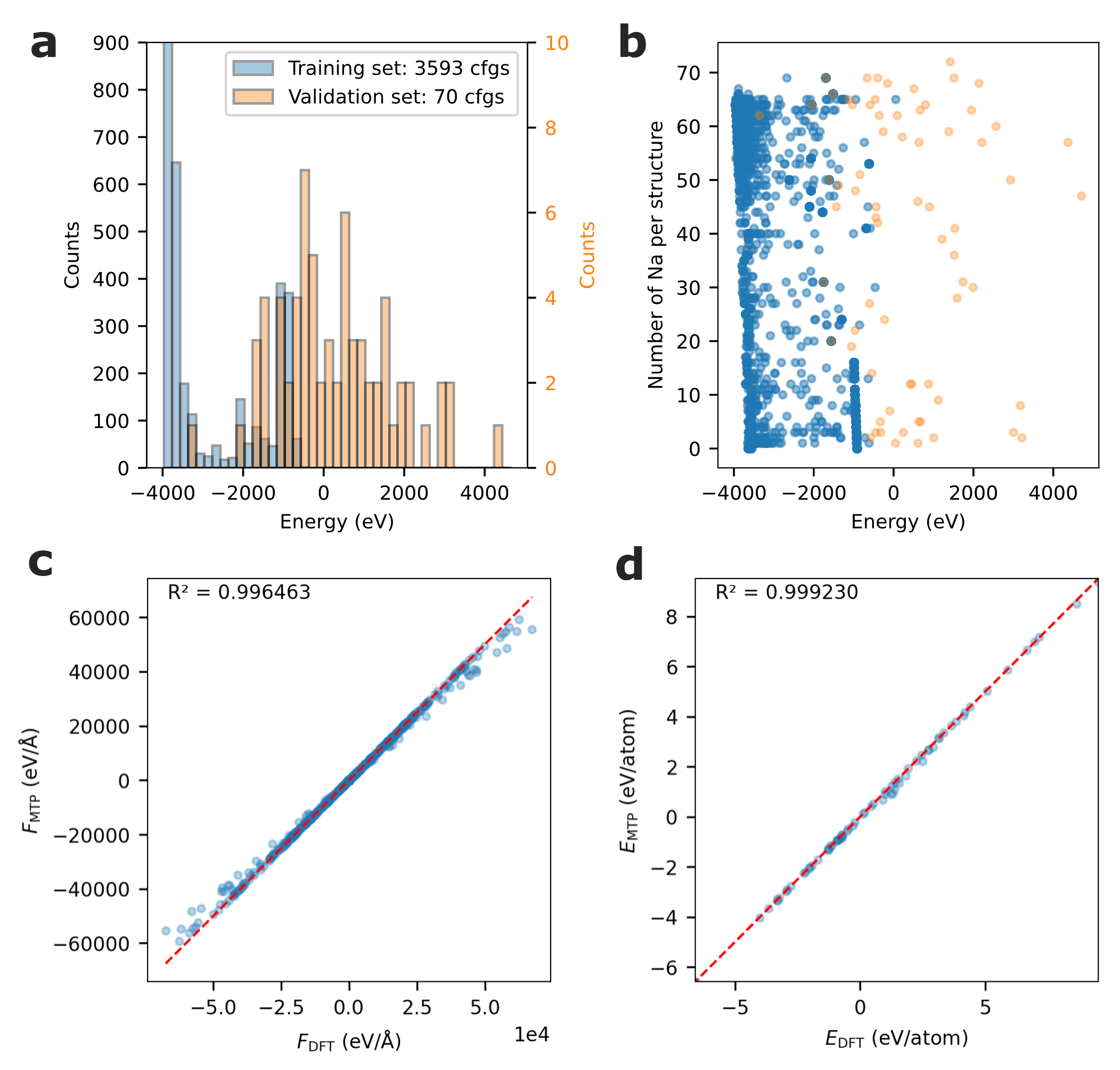}% Here is how to import EPS art
\caption{\label{fig:fig_S3}(a) and (b) The distribution of structural data within the training set (blue) and the validation set (orange), containing mostly structures of energies above 0 eV and interatomic forces exceeding 150 eV/Å, which is the screening criterion for the training set. Thus, the trained MTP can only predict the energies and forces through extrapolation. Plots of (c) MTP-predicted vs DFT forces and (d) MTP-predicted vs DFT energies prt atom. The trained MTP managed to reliably predict the energies and forces, proving the extrapolation capability and robustness.}
\end{figure*}

\begin{figure*}
\includegraphics{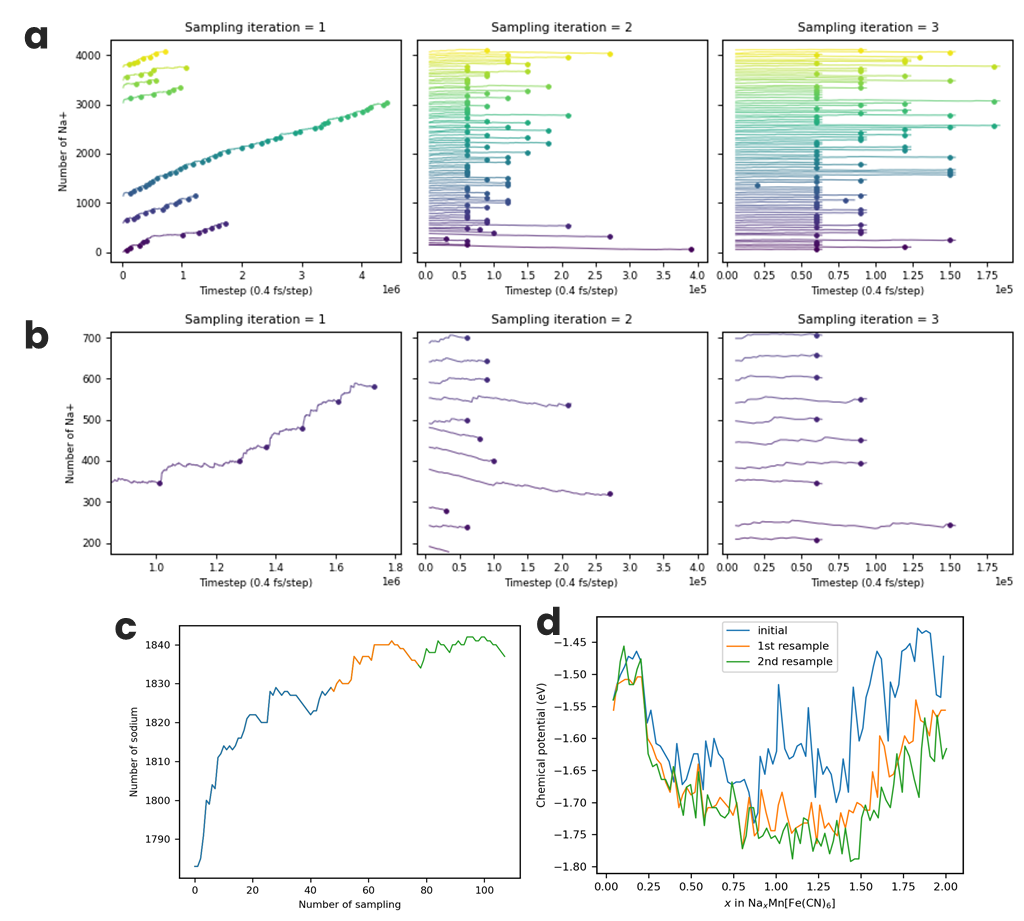}% Here is how to import EPS art
\caption{\label{fig:fig_S4}(a) and (b) Initial and first and second $\text{GCMC}$ resampling processes at each $N_{\text{Na}}^t$. The dots mark the convergence of the sampling. During the initial sampling, after the sodium concentration reaches equilibrium at each target value, a new bias potential centered at a different $N_{\text{Na}}^t$ is applied. First and second $\text{GCMC}$ resampling are conducted at each $N_{\text{Na}}^t$, respectively, to ensure convergence toward equilibrium. (c) Time evolution of sodium content during initial sampling (blue), first resampling (orange), and second resampling (green) at a $N_{\text{Na}}^t$. (d) Chemical potential profiles obtained from initial (blue), first (orange), and second (green) sampling rounds, illustrating the improvement in accuracy and reduction of noise with repeated sampling.}
\end{figure*}

\nocite{*}

\end{document}